\documentclass[aps,pra,twocolumn,superscriptaddress,amsmath,amssymb]{revtex4-2}

\usepackage{easyReview}

\usepackage{graphicx}
\usepackage{subfigure} 
\newcommand{\Eref}[1]{Eq. (\ref{#1})}
\newcommand{\Sref}[1]{Sec. \ref{#1}}
\newcommand{\Fref}[1]{Fig. \ref{#1}}

\begin{document}

\title{Resonance of Geometric Quantities and Hidden Symmetry in the Asymmetric Rabi Model}

\author{Qinjing Yu}
\affiliation{Key Laboratory of Artificial Structures and Quantum Control (Ministry of Education), School of Physics and Astronomy,\\ Shanghai Jiao Tong University, Shanghai 200240, China}

\author{Zhiguo L\"u}
\email[]{zglv@sjtu.edu.cn}
\affiliation{Key Laboratory of Artificial Structures and Quantum Control (Ministry of Education), School of Physics and Astronomy,\\ Shanghai Jiao Tong University, Shanghai 200240, China}
\affiliation{Collaborative Innovation Center of Advanced Microstructures, Nanjing University, Nanjing 210093, China}

\date{\today}

\begin{abstract}
We present the interesting resonance of two kinds of geometric quantities, namely the Aharonov-Anandan (AA) phase and the time-energy uncertainty, and reveal the relation between resonance and the hidden symmetry in the asymmetric Rabi model by numerical and analytical methods. By combining the counter-rotating hybridized rotating-wave method with time-dependent perturbation theory, we solve systematically the time evolution operator and then obtain the geometric phase of the Rabi model. In comparison with the numerically exact solutions, we find that the analytical results accurately describe the geometric quantities in a wide parameter space. We unveil the effect of the bias on the resonance of geometric quantities, (1) the positions of all harmonic resonances stemming from the shift of the Rabi frequency at the presence of the bias; (2) the occurrence of even order harmonic resonance due to the bias. When the driving frequency is equal to the subharmonics of the bias, the odd higher-order harmonic resonances disappear. Finally, the hidden symmetry has a resemblance to that of the quantum Rabi model with bias, which indicates the quasienergy spectra are similar to the energy spectra of the latter. 
\end{abstract}

\maketitle

\section{Introduction}\label{Sec introduction}
Since geometric phase of Berry's seminal work was introduced in the cyclic evolution of a system under adiabatic condition\cite{RN129}, its effects have been discovered and experimentally measured in various fields of physics\cite{RN115,RN150,RN151,RN152,RN153,RN283,RN307,RN362,add5,add6} as well as chemistry\cite{RN284}. The geometric property of Berry's phase visually lies in that it merely depends on the solid angle subtended by the closed path that the parameters traverse. Aharonov and Anandan (AA) extended Berry's phase to non-adiabatic cases\cite{RN128,RN137}, removing the adiabatic restriction and only assuming that the initial and final states of the quantum system differ by a total phase factor. Besides theoretical significance, AA phase has important applications to non-adiabatic geometric gates in quantum computation, with internal resiliency to certain noises and control errors\cite{RN305,RN331,RN361}. Furthermore, Samuel and Bhandari developed the geometric phase in a general setting without the assumption of cyclic evolution\cite{RN154}. All these phases are determined by the curve $ \mathcal{C} $ that the quantum state traverses in the projective
Hilbert space $ \mathcal{P} $, and are thus defined as geometric quantities\cite{RN136}.

Time-energy uncertainty, as one of the most fundamental quantities in quantum mechanics, is also proved to be a geometric quantity\cite{RN136,RN138}. Although the proposal of the uncertainty principle for time and energy even dates back to the origin of quantum mechanics in 1920s,  it is understood as setting a fundamental limit on the rate of quantum dynamics in its modern formulation\cite{RN286,RN291,RN290,RN292}, and thus widely employed in studies on the speed of gate operations in quantum computation\cite{RN295,RN294,RN297,RN296}, the precision of measurement in quantum metrology\cite{RN293,RN298,RN300,RN299} and so on. With the rapid development of technologies in manipulating quantum systems, it has received increasing attention as the basis for optimal control theory\cite{RN301,RN287,RN281,RN285}.

The semiclassical Rabi model \cite{RN261,RN262}, which describes a two-level system coupled with a classical monochromatic periodic field, is a typical quantum model exhibiting the properties of geometric quantities.  Its studies have a rich history for both experimental and theoretical investigations\cite{RN149,RN341,RN348,RN342,RN344,RN343,RN346,RN330,RN347}, and nowadays this model is widely applied in quantum information technology\cite{RN309,RN340,RN339}. It is found that there are a wide variety of interesting dynamical features in this model, such as well-known Rabi oscillations, Bloch-Siegert shifts, coherent destruction of tunneling (CDT)\cite{RN268}, driving-induced tunneling oscillations (DITO)\cite{RN269}, and plateau dynamics\cite{RN345}. In this work, we consider the asymmetric semiclassical Rabi model, where a bias term is taken into account.  The bias breaks the symmetry of the semiclassical Rabi model\cite{RN185}, which contributes to the complexity and phenomena different from those of the symmetric Rabi model. This allows for the representation of energy biases between the two states of a flux qubit\cite{RN258}.

While previous studies have primarily concentrated on the dynamics of the asymmetric semiclassical Rabi model\cite{RN147,RN103}, including phenomena such as CDT and DITO, relatively little attention has been paid to the geometric quantities of this model. The geometric quantities, which are robust against certain types of noise and errors, have the potential to be powerful tools for quantum computation and quantum information processing. Therefore, our work not only investigates the properties of geometric quantities of the asymmetric semiclassical Rabi model but also provides insights into the behavior in a wide range of parameter space.

In this work, we investigate the harmonic resonance of AA phase and the time-energy uncertainty of the asymmetric semiclassical Rabi model, which is beyond the work on the unbiased case\cite{RN102}. First, we generalize the harmonic resonance features of both geometric quantities in parameter space, and further explore the feature of the harmonic resonance. To derive the analytical expression for AA phase, we employ counterrotating-hybridized rotating-wave (CHRW) method, which takes account of the influences of counterrotating terms and bias together and is more reliable than rotating wave approximation. Further, considering the second harmonic terms of the Hamiltonian in the rotating frame, we combine perturbation theory with the CHRW method to calculate the AA phase and reveal its subtle resonant picture. Interestingly, at certain points in the parameter space, the odd harmonic resonance disappears in the numerical results. Using Floquet theory\cite{RN149,RN168}, we explain this interesting phenomenon, and reveal the relation between the hidden symmetry of the asymmetric semiclassical Rabi model and odd harmonic resonance. While the hidden symmetry of the asymmetric quantum Rabi model has been extensively studied, our results provide the first instructive evidence for the similar phenomenon in the semiclassical counterpart.

The structure of this work is as follows. In \Sref{Sec Resonance of Aharonov-Anandan phase and time-energy uncertainty}, after a brief review of both geometric quantities, we perform numerical calculations to demonstrate the resonance phenomenon, and show the features of harmonic resonance. In \Sref{Sec CHRW}, we employ the CHRW method to analytically calculate the AA phase and the positions of the higher-order harmonic resonances. In \Sref{Section Perturbation}, we apply perturbation theory based on the CHRW method to take into account the effects of higher-order harmonic terms in calculating the AA phase in the higher-order harmonic resonance regime. In \Sref{Sec Hidden symmetry}, we shed light on the absence of harmonic resonance in the results of numerical calculation as hidden symmetry appears. This is further discussed through comparison between the asymmetric semiclassical Rabi model and the asymmetric quantum Rabi model. Finally, we give the conclusion of this paper in \Sref{Sec Conclusion}.

	\section{Resonance of Aharonov-Anandan phase and time-energy uncertainty}\label{Sec Resonance of Aharonov-Anandan phase and time-energy uncertainty}
The Hamiltonian of the asymmetric semiclassical Rabi model reads
\begin{equation}\label{Eq classical Rabi H} 
	H(t)= -\frac{\Delta}{2} \sigma_x - \frac{\epsilon+A\cos(\omega t)}{2} \sigma_z,
\end{equation}
where $\sigma_x$ and $\sigma_z$ are the Pauli matrices, $\Delta$ is the tunneling strength, $\epsilon$ is the static bias. $ A, \omega $ and $T=2\pi/\omega$ are the amplitude, frequency and period of the driving field, respectively. We set $ \hbar = 1 $ throughout this paper and also use $ \varepsilon(t)=\epsilon+A\cos(\omega t) $ to denote the bias-modulated driving field\cite{RN147,RN343,add4}. This Hamiltonian can represent the systems moving in an effective double well potential modulated by an ac field, under the condition that only the ground state in either well can be occupied\cite{RN147,add3}. A system initially localized in one well will oscillate between the eigenstates in the left and right well due to quantum mechanical tunneling, which is exemplified by the two equivalent configurations of $ \mathrm{NH}_3 $, as well as the two current states of a flux qubit\cite{ RN258,add4}.

By implementing a rotation about the y-axis on this Hamiltonian, we obtain a transformed representation $ \exp (i\pi \sigma_y/4) H(t) \exp (-i\pi \sigma_y/4) =-\frac{\Delta}{2} \sigma_z + \frac{\epsilon+A\cos(\omega t)}{2} \sigma_x$. This particular form of the Hamiltonian is prevalently employed in the fields of quantum optics and nuclear magnetic resonance, where $ \Delta $ signifies the energy difference between the two levels, and the driving term causes transitions between these levels\cite{add1,add2}.

In the following, we first present a succinct overview of the concepts and geometric properties of the AA phase and the time-energy uncertainty in \Sref{Sec Aharonov-Anandan phase} and \Sref{Sec Time-energy uncertainty}. Then we display the $ \Delta $-dependence of both geometric quantities using numerical methods in \Sref{Sec Numerical calculations}, and further discuss the features of resonance in \Sref{Sec Features of resonance}.
\subsection{Aharonov-Anandan phase}\label{Sec Aharonov-Anandan phase}
Since the AA phase is a generalization of Berry phase, it is pertinent to briefly review the fundamental concepts and geometric properties of Berry's phase.

Consider a time-dependent Hamiltonian $H(\textbf{R}(t))$ where $\textbf{R}=(R_1, R_2, ..., R_n)$ and $\textbf{R}(T)=\textbf{R}(0)$, with the initial state being the $m$th eigenstate $|m(0)\rangle$ of $H(\textbf{R}(0))$.	If the system is subjected to adiabatic processes, i.e., $	|\langle j | \partial H/\partial t| k\rangle | \ll (E_j-E_k)^2, (j\ne k)$, where $E_j$ is the $j$th eigenenergy of $H(\textbf{R}(t))$, the state of the system will remain in the $m$th eigenstate $|m(t)\rangle$ while also obtaining a phase factor $\exp \left[ i (\alpha_m(t) +\gamma_m(t) \right]$, where $	\alpha_m(t)=- \int_0^t E_m(t')dt'$ is called the dynamical phase, and $	\gamma_m(t)=i\int_0^t \langle m(t')|\dot{m}(t')\rangle dt'$ is called the adiabatic phase. Berry‘s phase is defined as the adiabatic phase acquired over a cycle, i.e.,$ \gamma_m(T) $, which has been proved to be real and measurable.

The geometric property of Berry's phase is corresponding to  its relation to the solid angle subtended by the closed path in the parameter space spanned by $\textbf{R}$\cite{RN129}. Consequently, it is referred to as the geometric phase.

Based on Berry's phase, AA removed the adiabatic restriction, while assuming that the initial and final states of the quantum system differ by a total phase factor $e^{i\theta}$\cite{RN128}. Since the Hamiltonian in \Eref{Eq classical Rabi H} is periodic, the cyclic state $ |\psi(t)\rangle $ satisfies 
\begin{equation}
	|\psi(T)\rangle =e^{i\theta}|\psi(0)\rangle,
\end{equation} 
where $ \theta $ and $|\psi(0)\rangle$ are defined as the total phase and the cyclic initial state, respectively. Then the AA phase is defined by subtracting the dynamical phase $ \alpha $ from $ \theta $\cite{RN128}:
\begin{equation}\label{Eq definition for AA phase}
	\gamma=\theta-\alpha,
\end{equation}
where the dynamical phase is calculated by
\begin{equation}\label{Eq definition for dynamical phase}
	\alpha=-\int_{0}^{T}\langle\psi|H|\psi\rangle d\tau.
\end{equation}

The geometric property of AA phase is embodied in the projective Hilbert space $ \mathcal{P} $ (see Appendix \ref{App Projective Hilbert space}), in which the curve $ \mathcal{C} $ traversed by the system decides the value of AA phase. Quantities with such property are referred to as geometric quantities by AA.
\subsection{Time-energy uncertainty}\label{Sec Time-energy uncertainty}
Another geometric quantity under discussion is the time-energy uncertainty, defined as the time integral of the standard of energy:
\begin{equation}\label{Eq uncertainty}
	s=2\int \frac{\Delta E(t)}{\hbar}dt,
\end{equation}
where
\begin{equation}\label{Eq energy uncertainty}
	\Delta E(t)=\left[\langle \psi | H^2 |\psi \rangle-\langle \psi | H |\psi \rangle^2\right]^{1/2}.
\end{equation}
The coefficients in \Eref{Eq uncertainty} enables $ s $ to be equivalent to the distance of curve $ \mathcal{C} $ along which the state evolves in projective Hilbert space $ \mathcal{P} $, measured by the Fubini-Study metric\cite{RN136}. Owing to the periodicity of the Hamiltonian in \Eref{Eq classical Rabi H}, the curve $ \mathcal{C} $ will overlap beyond a single period. As such, it is adequate to compute the time-energy uncertainty within one period.

Suppose that $ |\psi\rangle $ and $ |\psi+d\psi\rangle $ are separated by an infinitesimal distance. Then the following expression for an infinitesimal length of path $ ds $ traversed by the state vector can be derived:
\begin{equation}\label{Eq ds}
	ds^2=4(1-|\langle \psi |\psi+d\psi \rangle^2|)
\end{equation}

The projective Hilbert space for a two-level system can be identified as the Bloch sphere. In the present case, the Fubini-Study metric is the usual metric on the Bloch sphere with unit radius. Therefore, $ ds $ can also be obtained as:
\begin{equation}\label{Eq d theta}
	ds=d\phi
\end{equation}
where $ \phi $ is the angle between $ |\psi\rangle $ and $ |\psi+d\psi\rangle $. The equivalence of \Eref{Eq ds} and \Eref{Eq d theta} can also be easily verified by means of numerical calculation.

\subsection{Numerical results}\label{Sec Numerical calculations}
\begin{figure*}[htbp]
	\centering
	\subfigure{
		\includegraphics[width=7.4cm]{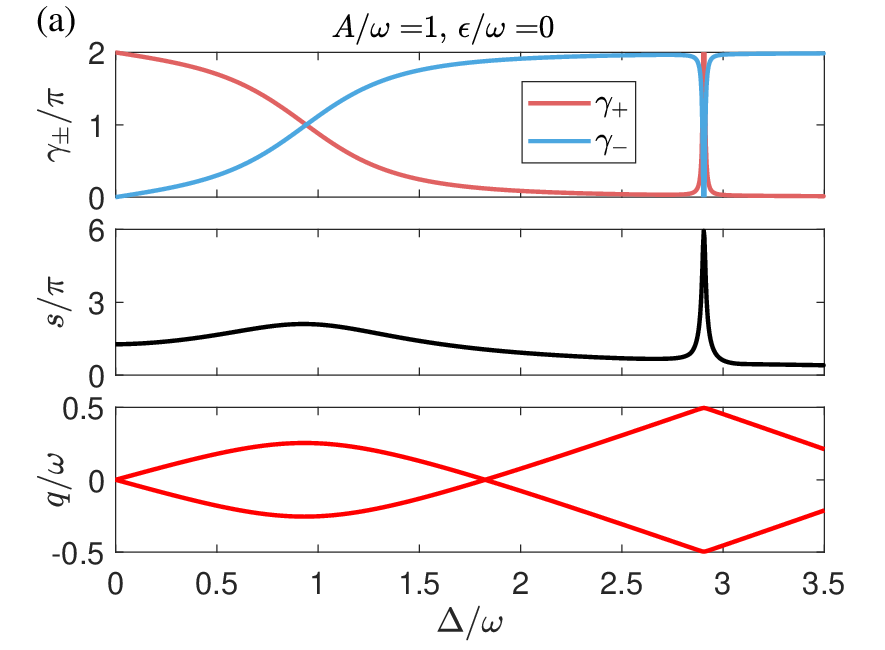}
		\label{Fig eps0}
	}
	\subfigure{
		\includegraphics[width=7.4cm]{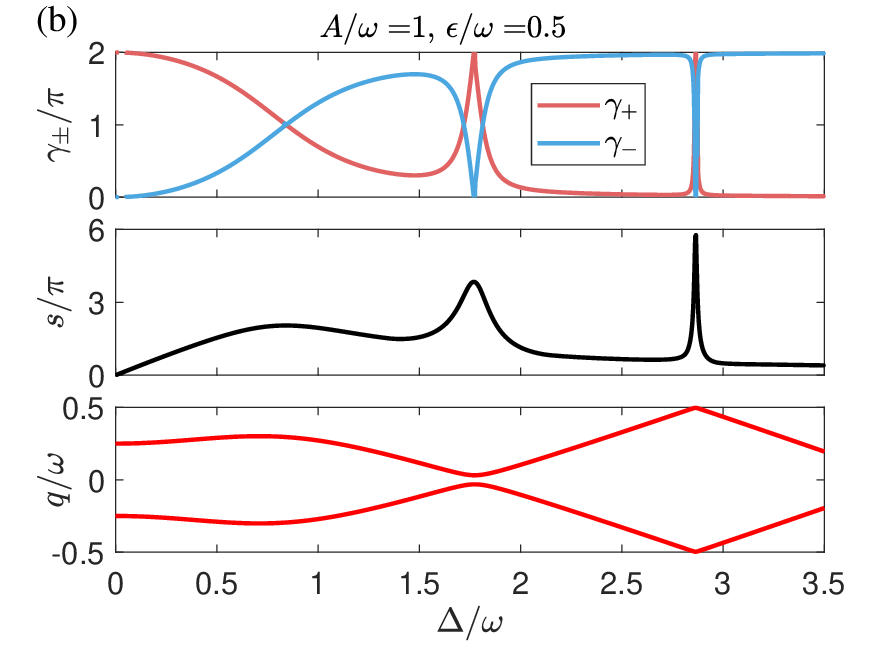}
		\label{Fig eps05}
	}
	\subfigure{
		\includegraphics[width=7.4cm]{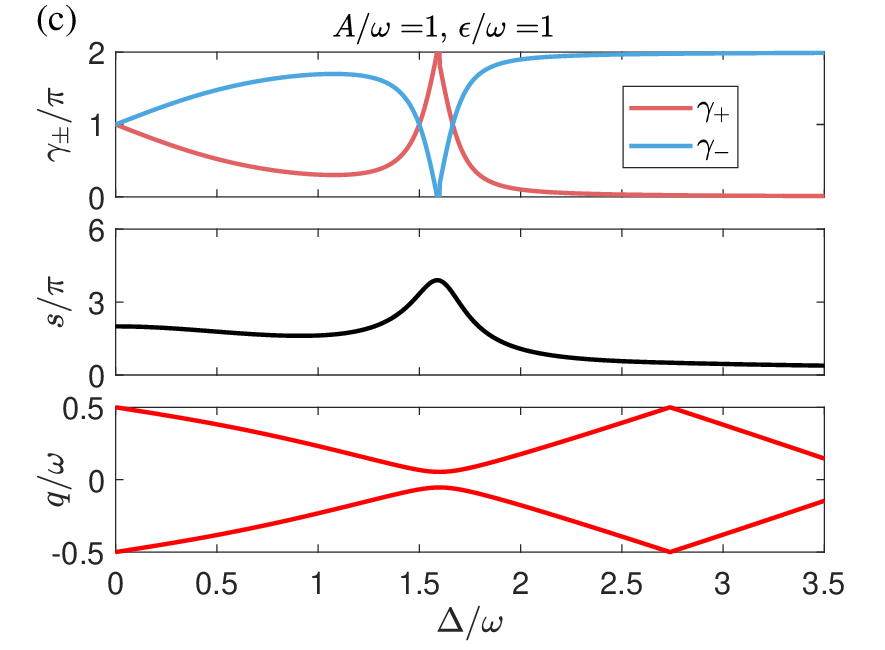}
		\label{Fig eps10}
	}
	\subfigure{
		\includegraphics[width=7.4cm]{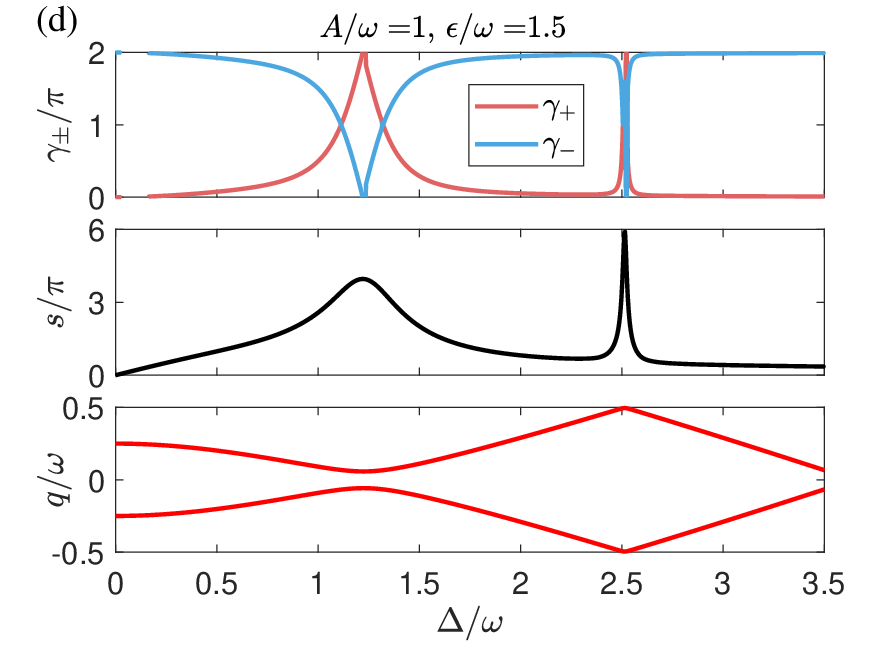}
		\label{Fig eps15}
	}
	\caption{AA phases $\gamma_{\pm}$, time-energy uncertainty $s$ and quasienergy $q$ (see Appendix \ref{App Floquet theory}) as a function of $\Delta/\omega$ for different values of $ \epsilon/\omega $ with a fixed driving strength $ A/\omega =1 $. (a) $ \epsilon/\omega=0$. (b) $\epsilon/\omega=0.5$. (c) $\epsilon/\omega=1$. (d) $\epsilon/\omega=1.5$. $ \gamma_+ $ and $ \gamma_- $ are AA phases corresponding to different cyclic initial states, and we set here $\gamma_-\equiv 2\pi -\gamma_+$ according to the complementary relation. $ \gamma_{\pm} $ and $ s $ are all divided by $\pi$, and $ q $ is divided by $ \omega $ in the figures. We have selected the interval $ [0,2\pi] $ as the principal value range for the AA phase.}
	\label{Fig AAandLength}
\end{figure*}

We present the $ \Delta $-dependence of the AA phase, the time-energy uncertainty and the quasienergy (see Appendix \ref{App Floquet theory}) for different bias using numerical methods in \Fref{Fig AAandLength}. With the Hamiltonian in \Eref{Eq classical Rabi H}, we first numerically solve the Schr\"odinger equation to obtain the evolution operator $ U(t) $:
\begin{equation}\label{Eq SchrodingerU}
	i\partial_t U(t) = H(t) U(t).
\end{equation}
Then both the total phase $\theta$ and the cyclic initial state $|\psi(0)\rangle$ can be obtained from the eigenvalues and eigenvectors of $ U(t=T) $. Finally, through numerical integration, we can calculate the AA phase using \Eref{Eq definition for AA phase} and \Eref{Eq definition for dynamical phase}, and the time-energy uncertainty using \Eref{Eq uncertainty} and \Eref{Eq energy uncertainty}. On the other hand, the quasienergy is obtained by diagonalizing the Floquet Hamiltonian numerically, which we will discuss in detail in \Sref{Sec Hidden symmetry}.

 For the semiclassical Rabi model, we obtain AA phases $\gamma_{\pm}$ which correspond to different cyclic initial states and satisfy $ \gamma_+ + \gamma_- = 2l \pi,l\in \mathbb{Z} $ (see Appendix \ref{App analysis of HS}). Nevertheless, the values of their time-energy uncertainty are equal, which means their trajectories in the projective Hilbert space are identical. Note that we have selected the interval $ [0,2\pi] $ as the principal value range for the AA phase and set $ \gamma_+ + \gamma_- = 2 \pi$. 

We show the characters of the resonance from the three quantities: AA phases $\gamma_{\pm}$, time-energy uncertainty $s$ and quasienergy $q$. The resonance feature of the time-energy uncertainty is quite pronounced. At the harmonic resonance, the uncertainty reaches its local maxima, which are close to $ 2\pi, 4\pi$ and $ 6\pi $ for the main, second and third resonance respectively. Meanwhile, in the vicinity of the harmonic resonance corresponding to the uncertainty, both branches of the AA phase $ \gamma_{\pm} $ pass through $ \gamma=\pi $, and the number of intersections increases with the order of resonance. At the same time, the branches of quasienergy exhibits avoided crossing. Through numerical calculations, we have found that the harmonic resonance of the three quantities tend to coincide.

In the top and middle panels of Fig. \ref{Fig eps0} we displays the geometric quantities for the Rabi model without bias, i.e., $ \epsilon=0 $. In this case, only the main and third harmonic resonance occur at $ \Delta/\omega \approx 1 $ and $ \Delta/\omega \approx 3 $, respectively. Near these resonances, $ \gamma_+ $ (or $ \gamma_- $) passes through $ \pi $ once and more than once, respectively. In the bottom panel of Fig. \ref{Fig eps0}, quasienergy spectrum near the main and third harmonic resonance regime  exhibits anti-crossing but that near $\Delta/\omega=2$ exhibits crossing.  It indicate that the corresponding relation between geometric quantities and quasienergy spectra.  

When $ \epsilon /\omega >0$, there happens the second harmonic resonance contributed by the bias term at  $\Delta/\omega\approx 2$ besides the main and third harmonic resonances, as observed in \Fref{Fig eps05}. With the increase of $ \epsilon /\omega $, all the resonance shift toward smaller values of $ \Delta/\omega $, and the main harmonic resonance eventually vanishes as $ \epsilon/\omega \ge 1 $, as shown in \Fref{Fig eps10} and \Fref{Fig eps15}. When $ \epsilon/\omega =1 $, the third harmonic resonance also disappears, because both geometric quantities are relevant to the hidden symmetry of the asymmetric semiclassical Rabi model, which we will come back to discuss in \Sref{Sec Hidden symmetry}.

\subsection{Features of resonance}\label{Sec Features of resonance}
\begin{figure*}[htbp]
	\centering
	\subfigure{
		\includegraphics[width=3.55cm]{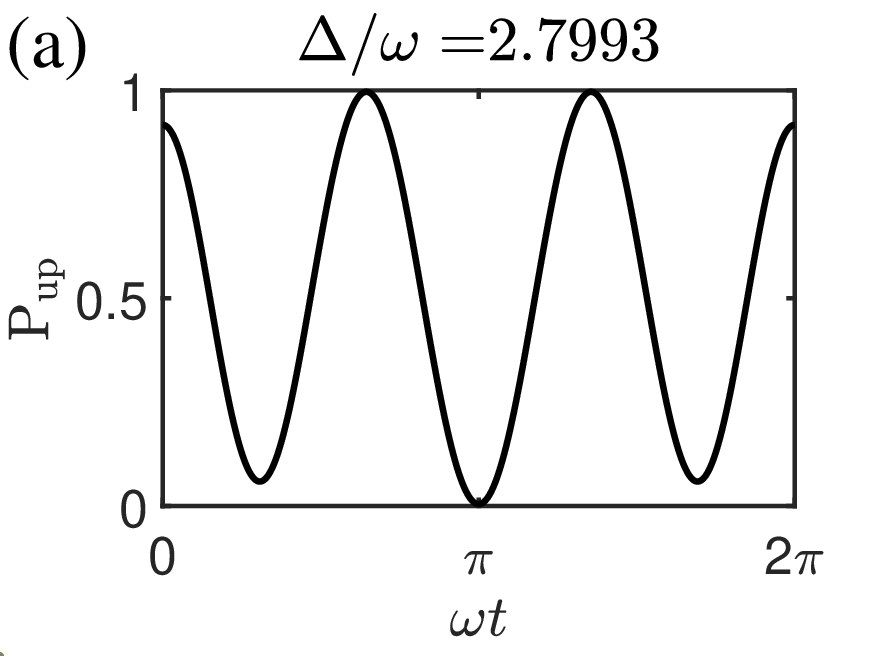}
		\label{Fig dy1}
	}
	\subfigure{
		\includegraphics[width=3.55cm]{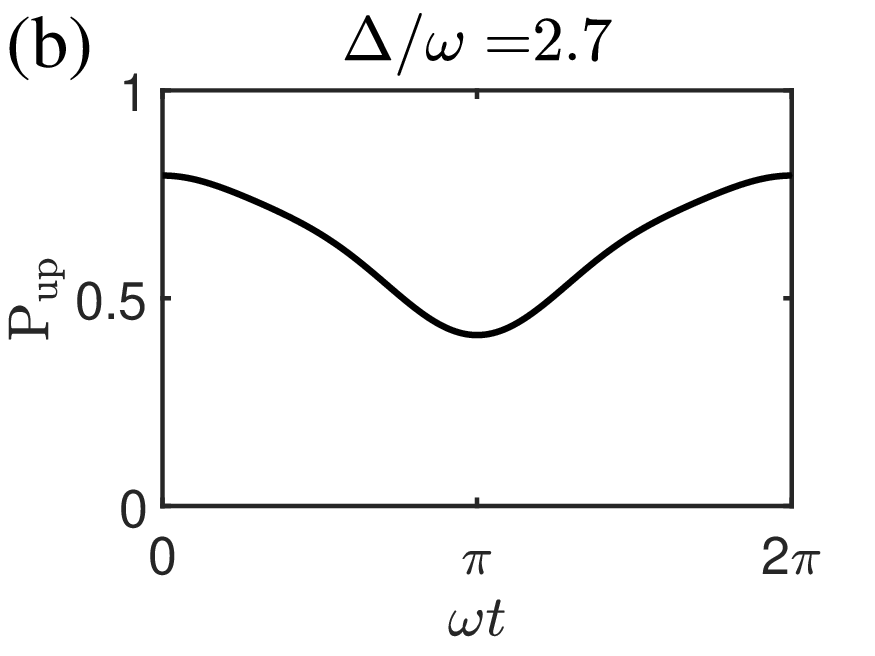}
		\label{Fig dy2}
	}
	\subfigure{
		\includegraphics[width=3.55cm]{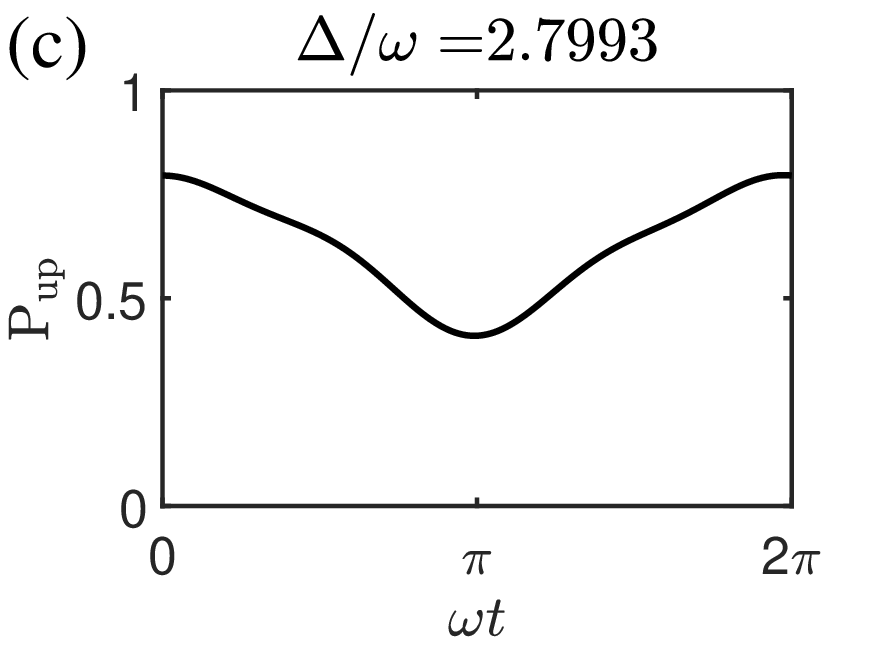}
		\label{Fig dy3}
	}
	\subfigure{
		\includegraphics[width=3.55cm]{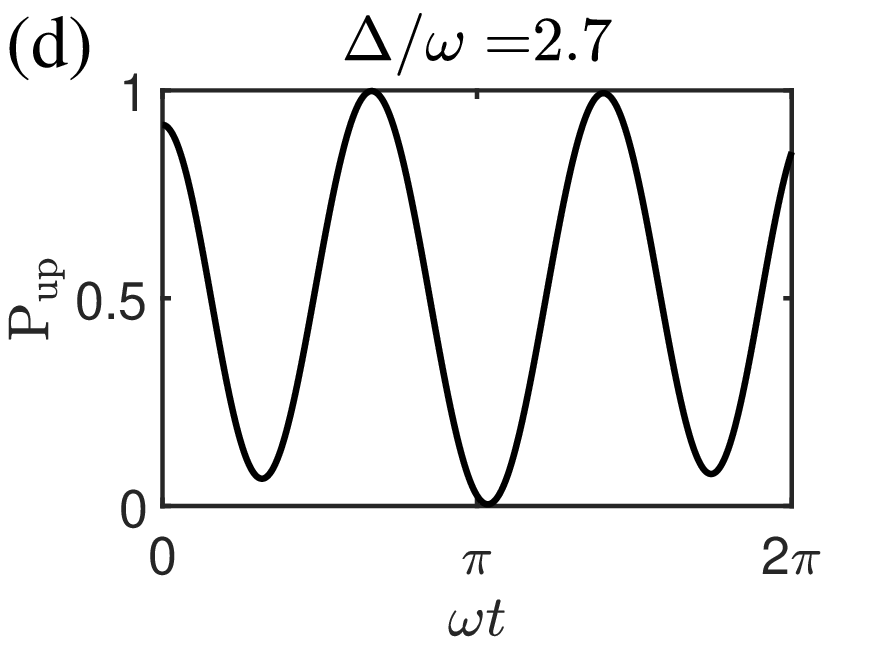}
		\label{Fig dy4}
	}
	\subfigure{
		\includegraphics[width=3.55cm]{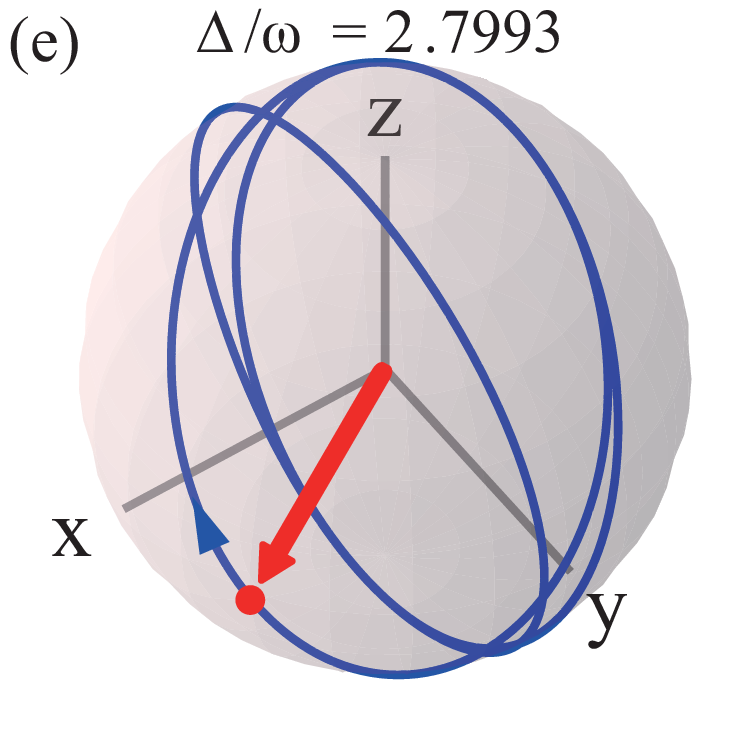}
		\label{Fig bloch1}
	}
	\subfigure{
		\includegraphics[width=3.55cm]{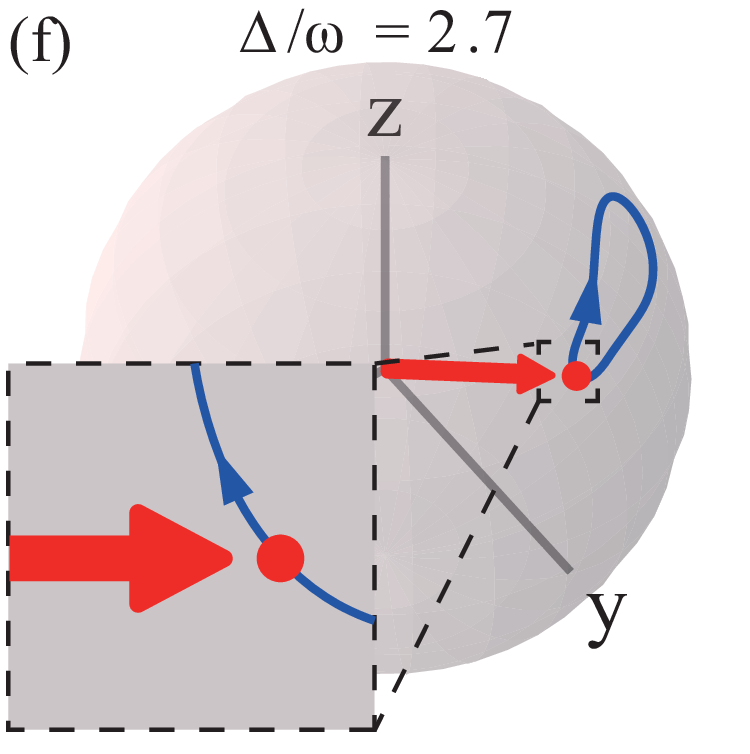}
		\label{Fig bloch2}
	}
	\subfigure{
		\includegraphics[width=3.55cm]{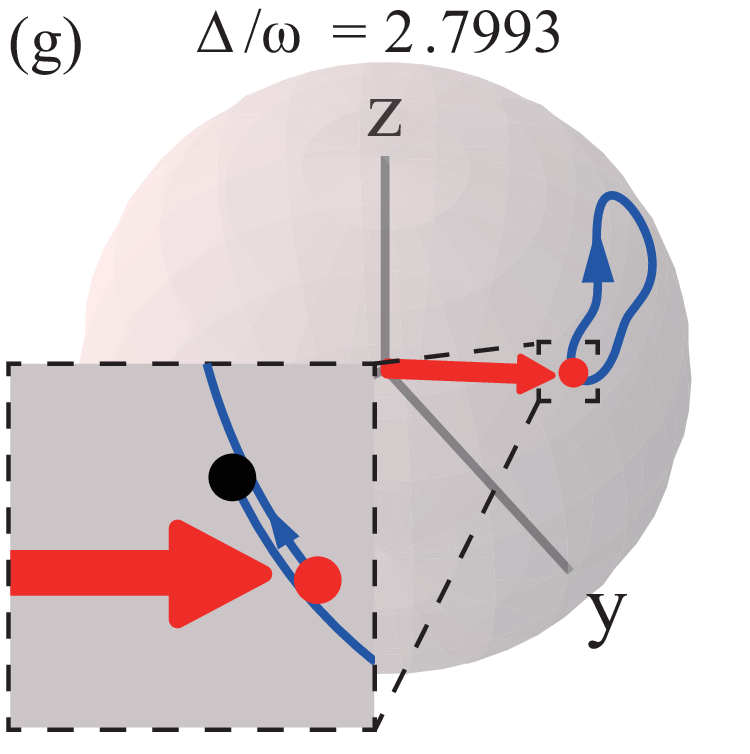}
		\label{Fig bloch3}
	}
	\subfigure{
		\includegraphics[width=3.55cm]{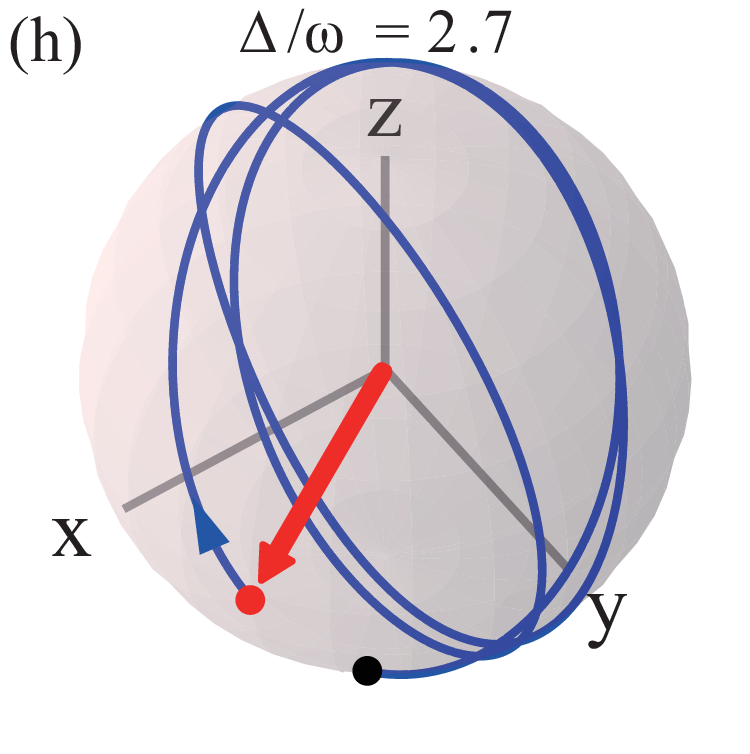}
		\label{Fig bloch4}
	}
	\caption{(a-d) Time evolution of $ P_{\rm{up}}$ for different tunneling strength $ \Delta/\omega $ and initial states, with $ A/\omega=1 $ and $ \epsilon/\omega= 0.8$ fixed: (a) $ \Delta/\omega=2.7993$ and the initial state is $|\psi_{\rm{res}}(0)\rangle$; (b) $\Delta/\omega=2.7$ and the initial state is $|\psi (0)\rangle|_{\Delta/\omega=2.7}$; (c) $\Delta/\omega=2.7993$ and the initial state is  $|\psi (0)\rangle|_{\Delta/\omega=2.7}$ ; (d)  $\Delta/\omega=2.7$ and the initial state is $|\psi_{\rm{res}}(0)\rangle$ . (e-h) Trajectories on the Bloch sphere with the parameters and the initial states corresponding to (a-d), respectively. The red arrows and dots on the Bloch spheres represent the initial states, while the black dots label the final states. The arrows on the trajectories represent the directions of quantum state evolution. In (f) and (g), we zoom in on the divergence and the convergence of the initial and the final states, respectively.}
	\label{Fig Bloch_sphere}
\end{figure*}

To explore the condition of resonance and visualize its features, in \Fref{Fig Bloch_sphere}, we demonstrate the population dynamics of the Rabi model and trajectories on the Bloch sphere for different tunneling strength $\Delta/\omega$ and initial states. For $ A/\omega=1 $ and $ \epsilon/\omega= 0.8$, the third harmonic resonance happens at $\Delta/\omega=2.7993$, and we use $|\psi (0)\rangle\equiv |\psi_{\rm{res}}(0)\rangle$ to denote one of its cyclic initial states, while at the near-resonance position $\Delta/\omega=2.7$, one of its cyclic initial states is denoted as $|\psi (0)\rangle|_{\Delta/\omega=2.7}$. Figures 2(a)-(d) show the population of the spin-up state $[1\quad0]^{T}$ in the $\sigma_z$ basis as a function of $ t$ for different cases, calculated by 
\begin{equation}
	{P_{\rm{up}}}( t)=\overline{ \frac{\sigma_z(  t)+1}{2}}=\left\langle \psi(0)\left|U(t)^\dagger\frac{\sigma_z+1}{2}U(t)\right|\psi(0)\right\rangle,
\end{equation} 
which represents the dynamics of the two-level systems. Figures 2(e)-(h) show the corresponding trajectories of $ |\psi(t)\rangle =U(t)| \psi(0)\rangle$ on the Bloch sphere, with red dots labeling the initial states and black dots labeling the final states.

In \Fref{Fig dy1} and \Fref{Fig bloch1},  $\Delta/\omega=2.7993$, which is the resonance position, and the initial state is $|\psi_{\rm{res}}(0)\rangle$. The features of resonance are manifestly shown by the rapid oscillation of $P_{\rm{up}}$ and the trajectory on the Bloch sphere whose length is very close to $6\pi$. As seen in \Fref{Fig bloch1}, the initial and final states converge because the initial state is the cyclic initial state for $\Delta/\omega=2.7993$. In \Fref{Fig dy2} and \Fref{Fig bloch2}, $\Delta/\omega=2.7$, which is corresponding to the nonresonance position, and the initial state is $|\psi (0)\rangle|_{\Delta/\omega=2.7}$. In this case, resonance is not observed as $P_{\rm{up}}$ varies slowly and the trajectory greatly contracts. 

In Figs. \ref{Fig dy3} and \ref{Fig bloch3}, we show the population dynamics and the trajectory on the Bloch sphere for  $\Delta/\omega=2.7993$ initially prepared from the initial cyclic state with $|\psi (0)\rangle|_{\Delta/\omega=2.7}$, respectively. We find that the dynamics and trajectory on the Bloch sphere in this case are similar to those results in \Fref{Fig dy2} and \Fref{Fig bloch2}, indicating that resonance disappears.	In \Fref{Fig dy4} and \Fref{Fig bloch4}, if  the initial state is initially prepared as $|\psi_{\rm{res}}(0)\rangle$ and a nonresonant parameter $\Delta/\omega=2.7$ is set, it is found that like-resonance phenomena recur which seems to those in \Fref{Fig dy1} and \Fref{Fig bloch1}. Note that the initial and final states diverge in \Fref{Fig bloch3} and \Fref{Fig bloch4} because in both cases the initial states and the parameters $\Delta/\omega$ do not match.

From \Fref{Fig Bloch_sphere}, we conclude that harmonic resonance mainly comes from the initial state $|\psi (0)\rangle$. As long as one cyclic initial state corresponding to its resonance position is selected, like-Resonance phenomena can happen even if the parameter $\Delta/\omega$ does not match that of the resonance. Meanwhile, $|\psi (0)\rangle$ is highly sensitive to $\Delta/\omega$ in the vicinity of harmonic resonance, as is observed from \Fref{Fig bloch1} and \Fref{Fig bloch3}.

\section{Counterrotating-hybridized	rotating-wave method}\label{Sec CHRW}
In this section, we apply the CHRW method to analytically calculate AA phase of the asymmetric semiclassical Rabi model. We first introduce the CHRW methodology in \Sref{Sec Methodology}, and then present the analytical expression for AA phase in \Sref{Sec Aharonov-Anandan phase 2}. Finally, we calculate the positions of the higher-order harmonic resonances by the CHRW method in \Sref{Sec Positions of the harmonic resonance}.

\subsection{Methodology}\label{Sec Methodology}
We perform the unitary transformation with a generator $S(t)=-i\frac{A}{2\omega}\sin(\omega t)\left(\xi \sigma_z+\zeta \sigma_x\right)$, to the Hamiltonian in Eq. (\ref{Eq classical Rabi H}). The two parameters $ \xi $ and $ \zeta $ are determined later.  In the rotating frame, we obtain the evolution operator
\begin{equation}
	i\frac{d {U'(t)}}{dt} =H'(t) U'(t)   
\end{equation}
in which $U'(t)=e^{S(t)}U(t)$ and $H'(t)=e^{S(t)}H(t)e^{-S(t)}-ie^{S(t)}\partial_t e^{-S(t)}$. Then, the wave function is obtained $|\psi'(t)\rangle=e^{S(t)} |\psi\rangle$, and the transformed Hamiltonian is written 
\begin{equation}
	\begin{split}
H'=&-\frac{\Delta}{2}\left[\sigma_x-\frac{1-\cos \Theta}{x^2} \xi\left(\xi \sigma_x-\zeta \sigma_z\right)+\frac{\sin \Theta}{x} \xi \sigma_y\right] \\
&-\frac{\varepsilon(t)}{2}\left[\sigma_z+\frac{1-\cos \Theta}{x^2} \zeta\left(\xi \sigma_x-\zeta \sigma_z\right)-\frac{\sin \Theta}{x} \zeta \sigma_y\right] \\
&+\frac{A}{2}(\xi \sigma_z+\zeta \sigma_x)\cos (\omega t),
	\end{split}
\end{equation}
where $\Theta=z \sin (\omega t)$, $z=\frac{A}{\omega}x$, and $x=\sqrt{\xi^2+\zeta^2}$.
After using the identity
\begin{eqnarray}
	\exp(iy \sin \alpha)=\sum_{-\infty}^{\infty}J_n(y)e^{in\alpha},
\end{eqnarray}
where $ J_n(y) $ are the nth-order Bessel functions of the first kind, we divide the Hamiltonian into four parts according to the order of the harmonics,
\begin{equation}
	H'=H_0'+H_1'(t)+H_2'(t)+V(t),
\end{equation}
where
\begin{equation}
	 H_0'=-\frac{\tilde{\Delta}}{2}\sigma_x-\frac{\tilde{\epsilon}}{2}\sigma_z,
\end{equation}
\begin{equation}
	\begin{split}
H_1^{\prime}=&-\frac{(\Delta \xi-\epsilon \zeta)}{x} J_1(z) \sin (\omega t) \sigma_y-\frac{A}{2}\left[1-\xi-\zeta^2 j_c\right]\\
& \cos (\omega t) \sigma_z +\frac{A}{2} \zeta\left[1-\xi j_c\right] \cos (\omega t) \sigma_x,
	\end{split}
\end{equation}
\begin{equation}
	\begin{split}
H_2^{\prime}=&\frac{A}{2} \frac{\zeta}{x} J_1(z) \sin (2 \omega t) \sigma_y \\
&-\frac{(\Delta \xi-\epsilon \zeta)}{x^2} J_2(z) \cos (2 \omega t)\left(\xi \sigma_x-\zeta \sigma_z\right),
	\end{split}
\end{equation}
\begin{equation}
\begin{split}
V=&\frac{A}{2} \frac{\zeta}{x^2} J_2(z) \cos (3 \omega t)\left(\xi \sigma_x-\zeta \sigma_z\right)\\
&-\frac{[\Delta \xi-\varepsilon(t) \zeta]}{x^2} \sum_{n=2}^{\infty}\bigl\{x J_{2 n-1}(z) \sin [(2 n-1) \omega t] \sigma_y \\
&+J_{2 n}(z) \cos (2 n \omega t)\left(\xi \sigma_x-\zeta \sigma_z\right)  \bigr\}.
\end{split}
\end{equation}
The parameters $ \tilde{\Delta},\tilde{\epsilon} $ and $ j_c $ are defined as
\begin{equation}
\tilde{\epsilon}=\epsilon+\frac{\zeta}{x^2}\left[1-J_0(z)\right](\Delta\xi-\epsilon\zeta),
\end{equation}
\begin{equation}
\tilde{\Delta}=\Delta-\frac{\xi}{x^2}\left[1-J_0(z)\right](\Delta\xi-\epsilon\zeta),
\end{equation}
\begin{equation}
j_c=\frac{1-J_0(z)-J_2(z)}{x^2},
\end{equation}
respectively. Note that the zero-$\omega$ Hamiltonian $H_0'$ consists of the renormalized tunneling term and renormalized bias one, single-$\omega$ Hamiltonian $H_1'$ corresponds to single-harmonic processes, and double-$\omega$ Hamiltonian $H_2'$ specifies second-order harmonic processes. Finally, $ V(t) $ includes all higher-order harmonic terms $ n\omega,n\geq 3 $.
Until now, no approximations are made, and the effects of the counterrotating terms and bias are all taken into account. Next, we keep all the zeroth and first harmonics of the transformed Hamiltonian $ (n\omega, n = 0,1) $ i.e., $H'\approx H_0'+H_1'$ and neglect the higher-order harmonic terms that involves all multi-$ \omega $ or multi-photon assisted transitions $ (n\omega, n = 2,3,4,...) $. The validity of the omission of $ H_2'+V $ depends on the effects of the higher-frequency driving terms whose contribution to the dynamics is not prominent except for the ultra-strong driving-strength case\cite{RN103}.

The Hamiltonian $H'_0$ is diagonalized by a unitary matrix $D=u\sigma_z-v\sigma_x$ with $ u=\sqrt{\frac{1}{2}-\frac{\tilde{\epsilon}}{2\tilde{\Xi}}} $ and $v=\sqrt{\frac{1}{2}+\frac{\tilde{\epsilon}}{2\tilde{\Xi}}}$ to the form
\begin{equation}
	\tilde{H}_0=\frac{\tilde{\Xi}}{2}\tau_z,    
\end{equation}
where $\tilde{\Xi}=\sqrt{\tilde{\Delta}^2+\tilde{\epsilon}^2} $ is the renormalized energy splitting,$ \tau_z $ is the z-component pseudo-spin operator in the energy eigenbasis.

The Hamiltonian $H'_1$
is then changed to :
\begin{equation}
	\begin{split}
\tilde{H}_1&=D^{\dagger}H_1'D\\
&= \frac{(\Delta \xi-\epsilon \zeta)}{x} J_1(z) \sin (\omega t) \tau_y \\
&+\frac{A}{2}\left[1-\xi-\zeta^2 j_c\right] \cos (\omega t) \left(\frac{\tilde{\epsilon}}{\widetilde{\Xi}} \tau_z+\frac{\tilde{\Delta}}{\widetilde{\Xi}} \tau_x\right)\\
&+\frac{A}{2} \zeta\left[1-\xi j_c\right] \cos (\omega t)\left(\frac{\tilde{\epsilon}}{\widetilde{\Xi}} \tau_x-\frac{\tilde{\Delta}}{\widetilde{\Xi}} \tau_z\right)
	\end{split}
\end{equation}
$\tau_x$ and $\tau_y$ are, respectively, the x-component and the
y-component spin operators in the energy eigenbasis. In order to make the driving interaction term $ D^{\dagger}H_1'D $ hold the RWA-like form, we choose the two proper parameters $ \xi $ and $ \zeta $ to satisfy the following two self-consistent equations:
\begin{equation}\label{xizeta1}
0=\frac{A}{2}\left[\frac{\tilde{\Delta}}{\tilde{\Xi}}\left(1-\xi-\zeta^2j_c\right)+\frac{\tilde{\epsilon}}{\tilde{\Xi}}\zeta(1-\xi j_c)\right]-\frac{\Delta \xi-\epsilon\zeta}{x}J_1(z),
\end{equation}
\begin{equation}\label{xizeta2}
	0=\tilde{\epsilon}(1-\xi-\zeta^2j_c)-\tilde{\Delta}\zeta(1-\xi j_c).
\end{equation}
Therefore, we obtain:
\begin{equation}
	\begin{split}
\tilde{H}=&D^{\dagger}(H_0'+H_1')D\\ &=\frac{\tilde{\Xi}}{2}\tau_z + \frac{\tilde{A}}{2} \left[ \tau_+ \exp (-i\omega t) + \tau_- \exp(i\omega t) \right],
	\end{split}
\end{equation}
where $\tau_\pm=(\tau_x\pm i\tau_y)/2$ and $ \tilde{A} $ is the renormalized amplitude of the driving field resulting from the combination of the counterrotating coupling and static bias,
\begin{eqnarray}
	\tilde{A}=\frac{\Delta \xi-\epsilon \zeta}{x}2J_1 \left(z\right).
\end{eqnarray}

\subsection{Aharonov-Anandan phase}\label{Sec Aharonov-Anandan phase 2}
The solutions to the Schrödinger equation corresponding to $\tilde{H}$ is denoted as $|\tilde{\psi}\rangle$, and satisfy $|\tilde{\psi}\rangle=D^{\dagger}|\psi'\rangle=D^{\dagger}e^S |\psi\rangle$.
Meanwhile, the evolution operator $\tilde{U}(t)$ corresponding to $\tilde{H}$ is solved analytically
\begin{equation}\label{Eq U_tilde}
\begin{split}
&\tilde{U}(t)=\\
&\left[
\begin{array}{cc}
	e^{-i \frac{\omega t}{2}}\left[\cos \left(\frac{\tilde{\Omega} t}{2}\right)-\frac{i \tilde{\delta}}{\tilde{\Omega}} \sin \left(\frac{\tilde{\Omega} t}{2}\right)\right] & -e^{-i \frac{\omega t}{2}} \frac{i \tilde{A}}{\tilde{\Omega}} \sin \left(\frac{\tilde{\Omega} t}{2}\right) \\
	-e^{i \frac{\omega t}{2}} \frac{i \tilde{A}}{\tilde{\Omega}} \sin \left(\frac{\tilde{\Omega} t}{2}\right) & e^{i \frac{\omega t}{2}}\left[\cos \left(\frac{\tilde{\Omega} t}{2}\right)+\frac{i \tilde{\delta}}{\tilde{\Omega}} \sin \left(\frac{\tilde{\Omega} t}{2}\right)\right]
\end{array}\right],
\end{split}
\end{equation}
where $ \tilde{\Omega} = \sqrt{\tilde{\delta}^2 + \tilde{A}^2 } $ and $ \tilde{\delta} = \tilde{\Xi} -\omega $ are, respectively, the modulated Rabi frequency and the renormalized detuning parameter given by the CHRW method.

By the analytical expression of transformed evolution operator $\tilde{U}(t)$, we calculate the cyclic states $|\psi_{\pm}(t)\rangle$ by
\begin{equation}
	|\psi_{\pm}(t)\rangle=e^{-S(t)}D |\tilde{\psi}_{\pm}(t)\rangle=e^{-S(t)}D\tilde{U}(t)|\tilde{\psi}_{\pm}(0)\rangle.
\end{equation}
and total phases $ \theta_{\pm} $ by their eigenvalues $e^{i\theta_{\pm}} $. The dynamical phases are obtained 
\begin{equation}\label{Eq alpha formula}
\begin{split}
\alpha_{\pm} &=- \int_{0}^{T} \langle \psi_{\pm}|H|\psi_{\pm} \rangle d\tau\\
&=- \int_{0}^{T} \langle \tilde{\psi}_{\pm}(0)|\tilde{U}^{\dagger}D^{\dagger}e^SHe^{-S}D\tilde{U}|\tilde{\psi}_{\pm}(0) \rangle d\tau.
\end{split}
\end{equation}
By $\tilde{U}(T)$ in \Eref{Eq U_tilde}, 
\begin{equation}
	\tilde{U}(T) = -\cos \left(\frac{\tilde{\Omega}T}{2}\right)I +i\frac{\tilde{A}}{\tilde{\Omega}}\sin \left(\frac{\tilde{\Omega}T}{2}\right)\tau_x + i\frac{\tilde{\delta}}{\tilde{\Omega}}\sin \left(\frac{\tilde{\Omega}T}{2}\right)\tau_z,
\end{equation}
we get its eigenvectors and phases of eigenvalues
\begin{equation}
	|\tilde{\psi}_{\pm}(0)\rangle=\sqrt{\frac{2 \tilde{\Omega}}{\tilde{\Omega} \mp \tilde{\delta}}}\left(\begin{array}{c}
		\frac{1}{2} \mp \frac{\tilde{\delta}}{2 \tilde{\Omega}} \\
		\mp \frac{\tilde{A}}{2 \tilde{\Omega}}
	\end{array}\right),
\end{equation}
\begin{equation}
	\theta_{\pm}=\pm \frac{\tilde{\Omega}-\omega}{2}T,
\end{equation}
respectively.
Dynamic phases read
\begin{equation}\label{Eq alpha_CHRW}
	\begin{split}
\alpha_{\pm}&=\pm \frac{T}{4\tilde{\Omega}} \Bigg\{ \epsilon\Bigg[ \frac{\tilde{\epsilon}}{\tilde{\Xi}}\tilde{\delta}\left(1+J_0(z)\right) +\frac{2\xi \zeta}{x^2} \frac{\tilde{\Delta}}{\tilde{\Xi}} \tilde{\delta} \left(1-J_0(z)\right)\\
& - \frac{2\zeta}{x}\tilde{A} J_1(z)+\frac{\xi^2-\zeta^2}{x^2}\frac{\tilde{\epsilon}}{\tilde{\Xi}}\tilde{\delta}\left(1-J_0(z)\right) \Bigg]\\
&+A\tilde{A} \Bigg[\frac{\xi^2}{x^2} \frac{\tilde{\Delta}}{\tilde{\Xi}} +\frac{\xi \zeta}{x^2} \frac{\tilde{\epsilon}}{\tilde{\Xi}}\left(\frac{2J_1(z)}{z}-1\right) +\frac{2\zeta^2}{x^2}\frac{\tilde{\Delta}}{\tilde{\Xi}}\frac{J_1(z)}{z}\Bigg]\\
&+2 \Delta \Bigg[\frac{\zeta^2}{x^2} \frac{\tilde{\Delta}}{\tilde{\Xi}}\tilde{\delta} + \frac{\xi^2}{x^2} \frac{\tilde{\Delta}}{\tilde{\Xi}} J_0(z)\tilde{\delta} +\frac{\xi\zeta}{x^2}\frac{\tilde{\epsilon}}{\tilde{\Xi}}\tilde{\delta}(1-J_0(z))  \\
&+\frac{\xi}{x}\tilde{A}J_1(z) \Bigg] \Bigg\}
	\end{split}
\end{equation}
Finally we obtain the analytical expression of AA phases
\begin{widetext}
\begin{equation}\label{Eq theta_CHRW}
	\begin{split}
&\gamma_{\pm} =\theta_{\pm} -\alpha_{\pm} \\
&=\pm\frac{T}{4\tilde{\Omega}} \Bigg\{
		2\tilde{\Omega}\left(\tilde{\Omega}-\omega\right) 
		-\epsilon\Bigg[ \frac{\tilde{\epsilon}}{\tilde{\Xi}}\tilde{\delta}\left(1+J_0(z)\right) +\frac{2\xi \zeta}{x^2} \frac{\tilde{\Delta}}{\tilde{\Xi}} \tilde{\delta} \left(1-J_0(z)\right) - \frac{2\zeta}{x}\tilde{A} J_1(z)+\frac{\xi^2-\zeta^2}{x^2}\frac{\tilde{\epsilon}}{\tilde{\Xi}}\tilde{\delta}\left(1-J_0(z)\right) \Bigg]
		-A\tilde{A} \\
		&\Bigg[\frac{\xi^2}{x^2} \frac{\tilde{\Delta}}{\tilde{\Xi}} +\frac{\xi \zeta}{x^2} \frac{\tilde{\epsilon}}{\tilde{\Xi}}\left(\frac{2J_1(z)}{z}-1\right) + \frac{2\zeta^2}{x^2}\frac{\tilde{\Delta}}{\tilde{\Xi}}\frac{J_1(z)}{z} \Bigg]-2 \Delta \left[ \frac{\zeta^2}{x^2} \frac{\tilde{\Delta}}{\tilde{\Xi}}\tilde{\delta} + \frac{\xi^2}{x^2} \frac{\tilde{\Delta}}{\tilde{\Xi}} J_0(z)\tilde{\delta} +\frac{\xi\zeta}{x^2}\frac{\tilde{\epsilon}}{\tilde{\Xi}}\tilde{\delta}(1-J_0(z)) \ +\frac{\xi}{x}\tilde{A}J_1(z) \right] 
		\Bigg\}.
	\end{split}
\end{equation}
\end{widetext}
For the symmetric case, i.e., $ \epsilon=0 $, the result of the parameter $ \zeta=0 $ is self-consistently solved by \Eref{xizeta1} and \Eref{xizeta2}. This simplifies \Eref{Eq theta_CHRW} to the previous result derived in \cite{RN102} albeit with slight differences in the coefficients due to the present definition of $\tilde{A}$:
\begin{equation}
	\gamma_\pm = \pm \left[ \frac{\tilde{\Omega}-\omega}{2}-\frac{\tilde{A}}{2\tilde{\Omega}}\left(\frac{\tilde{\Delta}\tilde{\delta}}{\tilde{A}}+\frac{A}{2}+\frac{\tilde{A}}{2}\right)\right]T.
\end{equation}
\subsection{Positions of the harmonic resonance}\label{Sec Positions of the harmonic resonance}
By the Rabi frequency of the CHRW method, we calculate the positions of the higher-order harmonic resonance\cite{RN102}. The second harmonic resonance occurs when the modulated effective Rabi frequency equals the frequency of external driving field, i.e.,
\begin{equation}\label{Eq condition_for_res2}
	\tilde{\Omega} = \omega.
\end{equation}
Likewise, the condition for the third harmonic resonance is
\begin{equation}\label{Eq condition_for_res3}
	\tilde{\Omega} = 2\omega.
\end{equation}
An analytical expression for the Rabi frequency up to the second order in the driving strength $ A $ is derived in Ref. \cite{RN103}:
\begin{equation}\label{Eq Rabi_freq}
	\tilde{\Omega}_{\rm 2nd}=\left(\omega-\Xi_0\right)^2+\frac{A^2\Delta^2}{2\Xi_0\left(\omega+\Xi_0\right)},
\end{equation}
where $ \Xi_0=\sqrt{\Delta^2+\epsilon^2} $. We use the condition $\tilde{\Omega}_{\rm 2nd}= \omega$ to estimate the position of the second harmonic resonance. 
\begin{figure}[htbp]
	\centering
	\subfigure{
		\includegraphics[width=7.5cm]{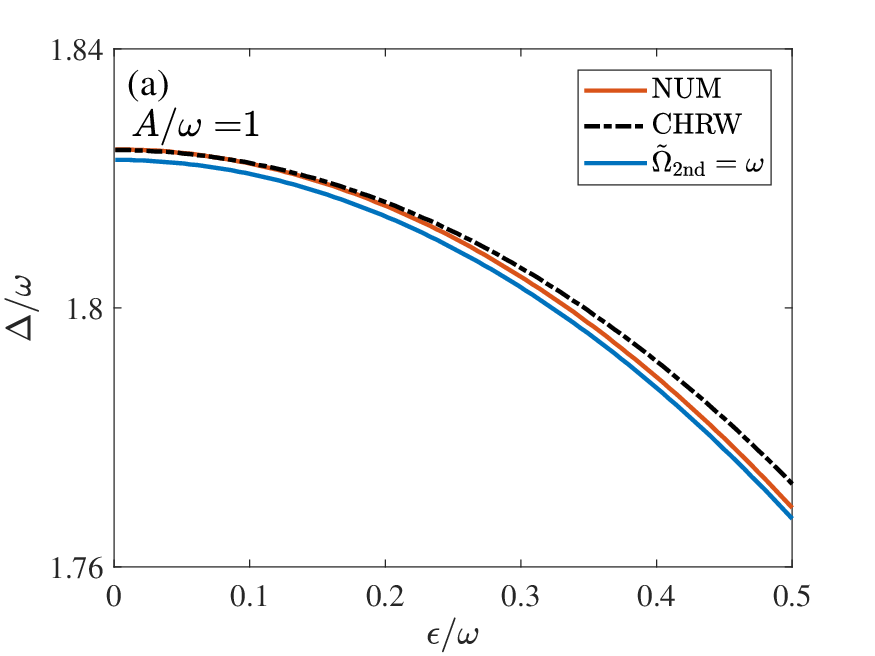}
		\label{Fig resonance peak2}
	}
	\subfigure{
		\includegraphics[width=7.5cm]{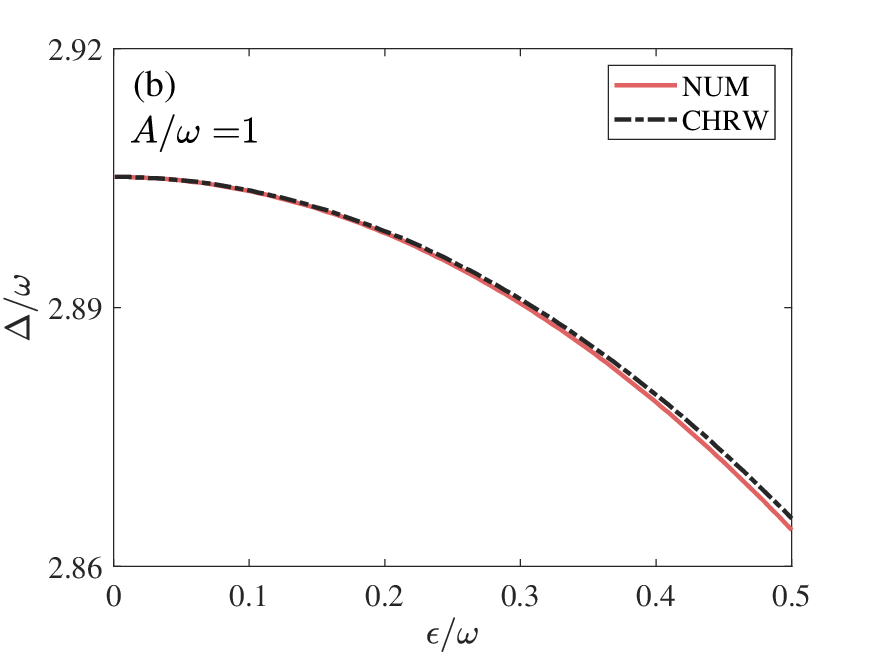}
		\label{Fig resonance peak3}
	}
	\caption{(a) Position of the second harmonic resonance. (b)  Position of the third harmonic resonance. $ 0<\epsilon/\omega<0.5, A/\omega=1 $ for both figures. The red lines are calculated by numerically exact method, the dashed dotted lines are obtained by \Eref{Eq condition_for_res2} and \Eref{Eq condition_for_res3} with the CHRW method, and the blue line in (a) is given by  $\tilde{\Omega}_{\rm 2nd}=\omega$ as an approximation.}
	\label{Fig resonance peak}
\end{figure}
It is obvious to see that, in \Fref{Fig resonance peak}, the positions of the harmonic resonance of numerically exact results agrees well with those of the CHRW method. Both the positions of the second harmonic resonance and third harmonic resonance show the tendency of decrease of the tunneling strength as the bias increases, indicating a compensating effect between $\Delta$ and $\epsilon$. The result obtained by using $\tilde{\Omega}=\omega$, which is shown by the blue line in \Fref{Fig resonance peak2}, is in agreement with the numerically exact solution when $\epsilon < \Delta$, despite a tiny deviation, which originates from neglecting higher order terms of $A$ in deriving \Eref{Eq Rabi_freq}. However, the relation $\tilde{\Omega}_{\rm 2nd}=2\omega$ could not correctly predict the third harmonic resonance because $\tilde{\Omega}_{\rm 2nd}$ does not take into account the competition effect of bias and counter-rotating terms in higher-order resonance parameter regime, such as $\Delta \sim 3 \omega$.     

\section{Perturbation theory}\label{Section Perturbation}
Although the CHRW method can predict the positions of the higher-order harmonic resonance of the AA phase and depict its asymptotic tendency as a function of $ \Delta/\omega $, it is necessary to combine it with perturbation theory in order to reveal the harmonic resonance features. Consider the time-dependent Hamiltonian $\tilde{H}'$ split into a unperturbed Hamiltonian $\tilde{H}$ and a perturbation with a dimensionless parameter $\lambda$ representing the order of perturbation,
\begin{equation}
	\tilde{H}'=\tilde{H}+\lambda\tilde{H}_{pt},
\end{equation}
and its corresponding time evolution operator
\begin{equation}
	\tilde{U}'=\tilde{U}+\lambda\tilde{U}_{pt}+...,
\end{equation}
where $ i\frac{d\tilde{U}}{dt} =\tilde{H}\tilde{U}$. Since $ \tilde{U}'(t) $ satisfies the time-dependent Schr\"odinger equation $ i\frac{d\tilde{U}'}{dt} =\tilde{H}'\tilde{U}' $, we can solve $ \tilde{U}_{pt} $ to the first order
\begin{equation}
	i\frac{d\tilde{U}_{pt}}{dt} = \tilde{H}\tilde{U}_{pt}+\tilde{H}_{pt}\tilde{U}.
\end{equation}
Then we analytically solve
\begin{equation}\label{Eq U_pt}
	\tilde{U}_{pt}(t) = -i\tilde{U}\int_0^t \tilde{U}^{-1}\tilde{H}_{pt}\tilde{U} d\tau.
\end{equation}
In this work we choose the parameter, $ A/\omega <2 $, and therefore the influences of second harmonic terms in $ H_2'(t) $ are dominant compared with the other higher harmonic terms in $ V(t) $. In the following, by considering the second harmonic terms in $ H_2'$, we calculate evolution operator under perturbation and then derive modified cyclic initial states and more accurate analytical results for the AA phase. In the energy eigenbasis representation of diagonalization, $ H_2' $ becomes  
\begin{equation}
	\tilde{H}_2=D^{\dagger}H_{2}'D=\tilde{H}_{2y}+\tilde{H}_{2x}+\tilde{H}_{2z},
\end{equation}
where
\begin{equation}
	\tilde{H}_{2y}=-\frac{A\zeta}{2x}J_1(z)\sin (2\omega t)\tau_y,
\end{equation}
\begin{equation}
	\tilde{H}_{2x}=\frac{\epsilon \zeta-\Delta \xi}{x^2}J_2(z) q \cos (2\omega t) \tau_x, 
\end{equation}
\begin{equation}
	\tilde{H}_{2z}=\frac{\epsilon \zeta-\Delta \xi}{x^2}J_2(z) p \cos (2\omega t) \tau_z,
\end{equation}
and $p = (v^2-u^2)\zeta-2\xi uv$ and $ q =(v^2-u^2)\xi +2uv\zeta$. Here we take into consideration the effects of the terms with $ \tau_x,\tau_y $ and $ \tau_z $ separately.
First, we explore the effects of $ \tilde{H}_{2y} $, calculating the integral:
\begin{equation}
\begin{split}
&\int_{0}^{T} \tilde{U}^{-1}\tilde{H}_{2y}\tilde{U} d\tau = \frac{2A\zeta J_1(z) \sin \left(\frac{\tilde{\Omega}}{2}T\right)}{x\tilde{\Omega}^2 (9\omega^4 -10\omega^2\tilde{\Omega}^2 +\tilde{\Omega}^4)} \\
&\Bigg[ \left( 3\tilde{\delta}\tilde{\Omega}\omega^3+ 2\tilde{\delta}^2\tilde{\Omega}\omega^2-\tilde{\delta}\tilde{\Omega}^3\omega \right) \cos\left(\frac{\tilde{\Omega}}{2}T\right) \tau_x \\
&+ \left( \tilde{\Omega}^4\omega-2\tilde{\Omega}^2\tilde{\delta}\omega^2-3\tilde{\Omega}^2\omega^3 \right)\sin \left(\frac{\tilde{\Omega}}{2}T\right) \tau_y \\
&+ \tilde{A} \left(\tilde{\Omega}^3 \omega -3\tilde{\Omega}\omega^3-2\tilde{\Omega}\tilde{\delta}\omega^2 \right)\cos\left(\frac{\tilde{\Omega}}{2}T\right) \tau_z 	\Bigg]. 
\end{split}
\end{equation}
Substituting \Eref{Eq U_tilde} into \Eref{Eq U_pt}, we obtain its first order correction to $ \tilde{U}' $:
\begin{equation}
	\tilde{U}_{2y}(T)=i \frac{2A\zeta \omega J_1(z) \sin (\frac{	\tilde{\Omega}}{2}T) (3\omega+2\tilde{\delta}\omega-\tilde{\Omega}^2)  }{x\tilde{\Omega}(9\omega^4 -10\omega^2\tilde{\Omega}^2 +\tilde{\Omega}^4)}(\tilde{\delta}\tau_x -\tilde{A}\tau_z).
\end{equation}
The whole time evolution operator $ \tilde{U}'_{y}(T) = \tilde{U}(T) + \tilde{U}_{2y}(T) $ yields
\begin{equation}
\begin{split}
\tilde{U}'_{y}(T)&=-\cos \left(\frac{\tilde{\Omega}T}{2}\right)I +\frac{i}{\tilde{\Omega}}\sin \left(\frac{\tilde{\Omega}T}{2}\right)\\
&\left[ (\tilde{A}+{k_y}\tilde{\delta}) \tau_x+(\tilde{\delta}-{k_y}\tilde{A} )\tau_z  \right],
\end{split}
\end{equation}
where 
\begin{equation}\label{Eq ky}
	{k_y}=\frac{2A\zeta \omega J_1(z) (3\omega+2\tilde{\delta}\omega-\tilde{\Omega}^2)  }{x(9\omega^4 -10\omega^2\tilde{\Omega}^2 +\tilde{\Omega}^4)}.
\end{equation}  
We immediately obtain one of its eigenvectors and its corresponding eigenvalue
\begin{equation}\label{Eq tilde p y'}
	|\tilde{\psi}_{+y}'(0) \rangle= \frac{1}{{L_y}}\left[
	\begin{array}{c}
		\tilde{A}+{k_y}\tilde{\delta} \\
		{k_y}\tilde{A}-\tilde{\delta}-\sqrt{1+{k_y}^2}\tilde{\Omega}
	\end{array} \right],
\end{equation}
\begin{equation}\label{Eq eigenvalue_pt}
	\lambda_+ = -\cos \left(\frac{\tilde{\Omega}T}{2}\right) -i\sqrt{1+{k_y}^2} \sin \left(\frac{\tilde{\Omega}T}{2}\right),
\end{equation}
where the normalization factor
\begin{equation}\label{Eq Ly}
	{L_y}=\left[2(1+{k_y}^2)\tilde{\Omega}^2-2(\tilde{\delta}+{k_y}\tilde{A})\sqrt{1+{k_y}^2}\tilde{\Omega}\right]^{1/2}.
\end{equation}
Because the eigenvalue $|\lambda_+|\simeq 1 $, we can regard
\begin{equation}\label{Eq total_phase pt}
	\theta_+' = {\rm{arg}}(\lambda_+)=\arctan \left[\sqrt{1+k_y^2} \tan \left(\frac{\tilde{\Omega}T}{2}\right)\right]-\frac{\omega T}{2}
\end{equation}
as the modified total phase under perturbation, which includes the effects of counterrotating terms. Also, the perturbed Rabi frequency in the vicinity of harmonic resonance can be derived accordingly (see Appendix \ref{App pt}). In fact, $ \theta_+' $ is very close to $\theta_+ $ given by the CHRW method, as the eigenvalues of $\tilde{U}'_{y}(T)$ are very close to those of $\tilde{U}(T)$. Thus we can use the latter to approximate $\theta_+' $, and the less $|{k_y}|$ is, the better the approximation is.

\begin{figure*}[htbp]
	\centering
	\subfigure{
		\includegraphics[width=7.5cm]{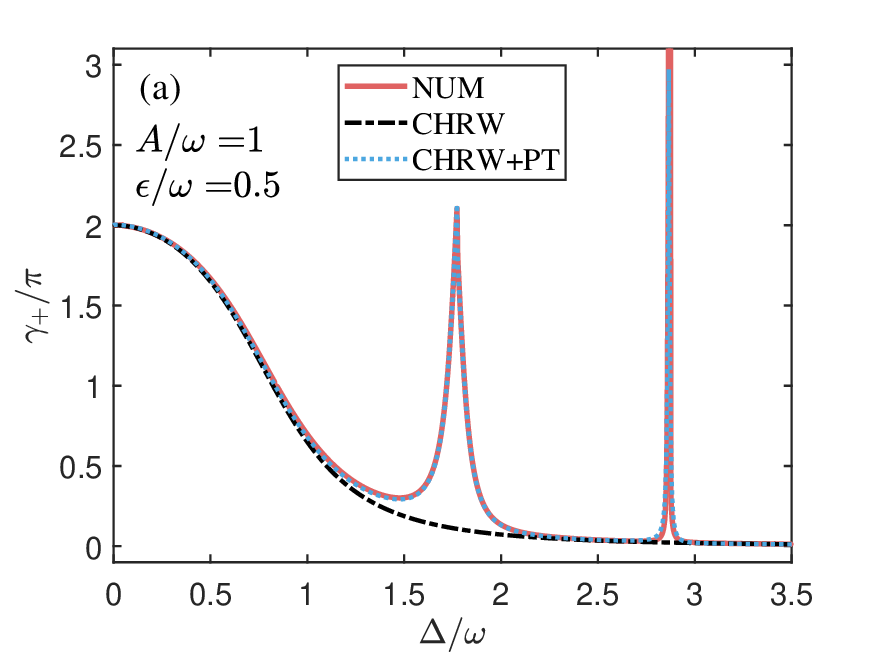}
		\label{Fig AAe05}
	}
	\subfigure{
		\includegraphics[width=7.5cm]{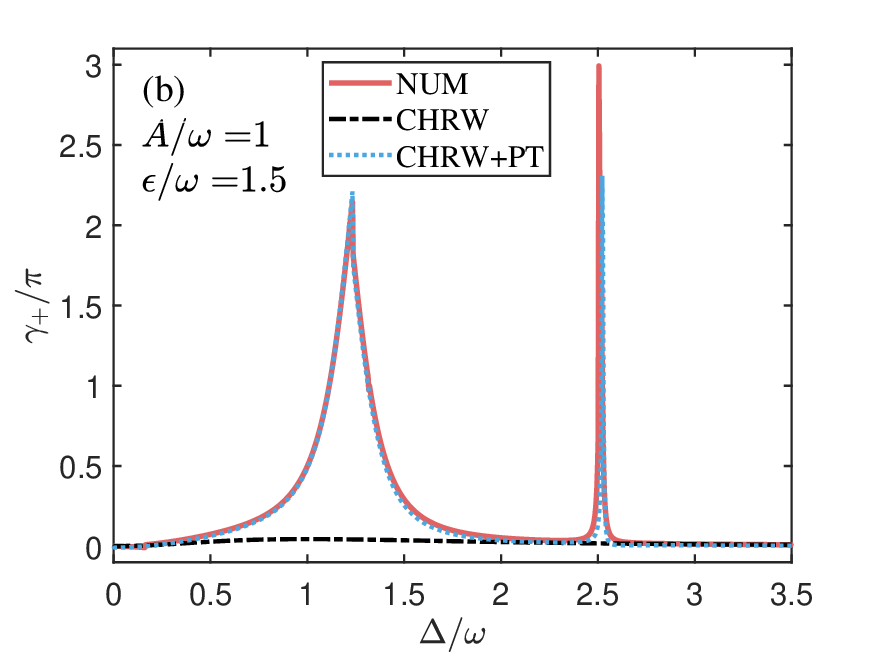}
		\label{Fig AAe15}
	}
	\subfigure{
		\includegraphics[width=7.5cm]{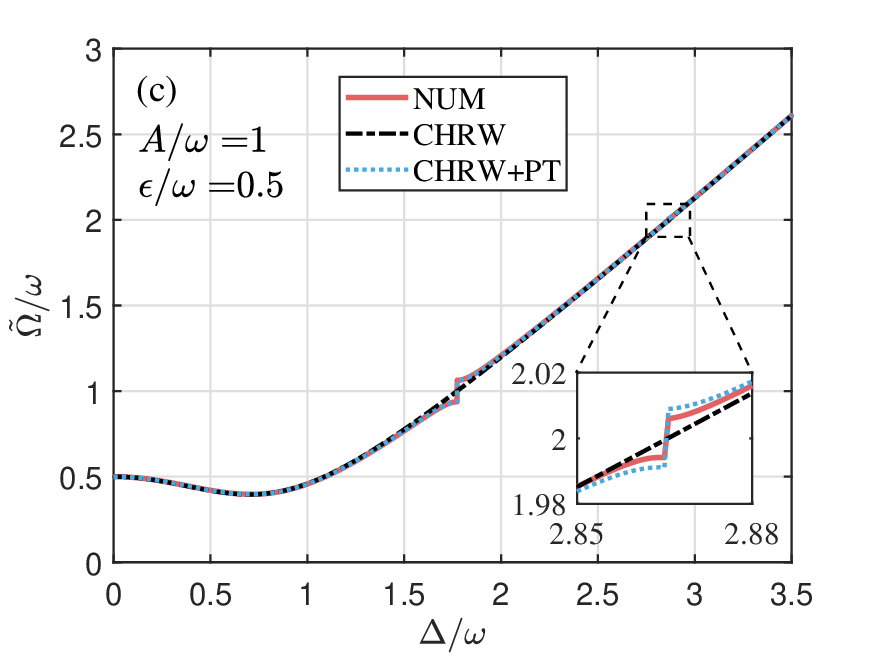}
		\label{Fig Omega_e05}
	}
	\subfigure{
		\includegraphics[width=7.5cm]{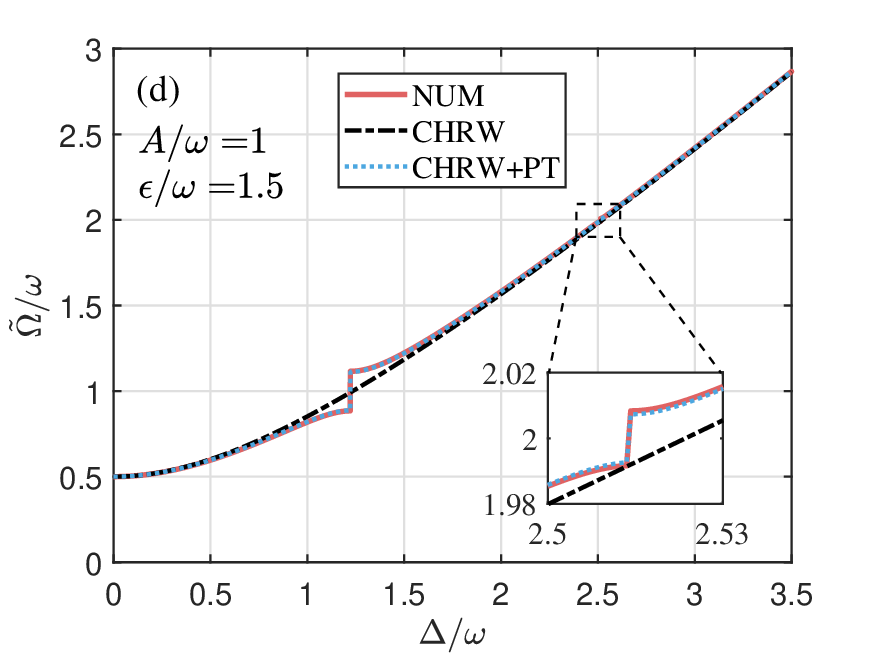}
		\label{Fig Omega_e15}
	}
	\caption{(a-b) AA phase $ \gamma_+/\pi $ as a function of $\Delta/\omega$ for different values of $ \epsilon/\omega $ with $A/\omega=1$ fixed: (a) $ \epsilon/\omega=0.5$; (b) $\epsilon/\omega=1.5$. (c-d) Modulated effective Rabi frequency $ \tilde{\Omega}/\omega $ as a function of $\Delta/\omega$ for different values of $ \epsilon/\omega $ with $A/\omega=1$ fixed: (c) $ \epsilon/\omega=0.5$; (d) $\epsilon/\omega=1.5$. The results of  perturbation theory based on the CHRW method, denoted as CHRW+PT, agree well with the numerically exact results, while those obtained by CHRW show asymptotic tendency. In (c) and (d), we include magnified views to highlight the sudden changes at the harmonic resonance at $\tilde{\Omega}/\omega\approx2$. }
	\label{Fig AA phase}
\end{figure*}

Owing to the great contribution $ \tilde{U}_{2y}(T) $ to $ \tilde{U}'_{y}(T) $, both dynamical phase and AA phase change sharply near higher-order harmonic resonance. Using \Eref{Eq definition for AA phase} and \Eref{Eq alpha formula}, we obtain the AA phase corresponding to \Eref{Eq tilde p y'}
\begin{equation}
	\gamma_+=\theta_+ +\int_{0}^{T} \langle \tilde{\psi}_{+y}'(0)|\tilde{U}^{\dagger}D^{\dagger}e^SHe^{-S}D\tilde{U}|\tilde{\psi}_{+y}'(0)\rangle d\tau
\end{equation}
Here we use $ \tilde{U} $ instead of $ \tilde{U}' $ in the integrand, which causes neglectable difference in the calculation. In light of the complementary relation $ \gamma_+ +\gamma_- = 2l\pi,l\in \mathbb{Z}$, we obtain both AA phases approximately
\begin{widetext}
\begin{equation}\label{Eq gamma_pt_2y}
	\begin{split}
\gamma_\pm\approx &\pm\frac{\tilde{\Omega}-\omega}{2}T\mp\frac{\pi}{2{L_y}^2} \Bigg\{ (\tilde{\delta}+\sqrt{{k_y}^2+1}\tilde{\Omega}-{k_y}\tilde{A}) \Big\{\frac{A^2 \tilde{\delta}}{2\omega^3}\left[\epsilon(\nu^2-\mu^2)-2\Delta \mu \nu\right]		-\frac{\tilde{\delta}}{128\omega}(3z^4-64z^2+512)\\
&\left(u^2\epsilon-2uv\Delta-v^2\epsilon\right)-\frac{2A\tilde{A}}{\omega^2}\left[(u\epsilon-v\Delta)\nu-(u\Delta+v\epsilon)\mu\right]+\frac{A\tilde{A}}{\omega}\left[\frac{2}{z}J_1(z)+1\right]\left[{k_y}(u^2-v^2)+2uv\right]\Big\}\\
&+\frac{4{k_y}}{x^2}J_2(z)\left(u\nu-v\mu\right)\left[{k_y}\tilde{A}(\tilde{\Omega}+\tilde{\delta})-\left(1+\sqrt{{k_y}^2+1}\right)\tilde{\Omega}\tilde{\delta}-\sqrt{{k_y}^2+1}\tilde{\Omega}^2-\tilde{\delta}^2\right]\Bigg\},
	\end{split}
\end{equation}

where $\mu = \xi u-\zeta v, \nu = \xi v+\zeta u$. The details of derivation are given in Appendix \ref{App pt}.

Likewise, when we take account of the influences of $ \tilde{H}_{2x} $ and $ \tilde{H}_{2z} $, we repeat the processes of derivation above and obtain
\begin{eqnarray}
	\tilde{U}_{2x}(T)=\frac{ik_x\sin \left(\frac{\tilde{\Omega}}{2}T\right)}{\tilde{\Omega}}\left(\tilde{\delta}\tau_x -\tilde{A}\tau_z\right), \\
	\tilde{U}_{2z}(T)=\frac{ik_z\sin \left(\frac{\tilde{\Omega}}{2}T\right)}{\tilde{\Omega}}\left(\tilde{\delta}\tau_x -\tilde{A}\tau_z\right). 
\end{eqnarray}
where
\begin{equation}\label{Eq kx}
k_x=\frac{2(\Delta\xi -\epsilon \zeta)J_2(z)}{x^2}
\frac{\left[(v^2-u^2)\xi +2uv\zeta\right](3\omega^3+5\omega^2\tilde{\delta}+\omega\tilde{\Omega}^2-\tilde{\delta}\tilde{\Omega}^2)}{(9\omega^4-10\omega^2\tilde{\Omega}^2+\tilde{\Omega}^4)},
\end{equation}
\begin{equation} \label{Eq kz}
k_z=-\frac{2(\Delta\xi -\epsilon \zeta)J_2(z)}{x^2}\frac{\tilde{A}\left[2\xi uv+\zeta (u^2-v^2)\right]}{\tilde{\Omega}^2-4\omega^2}.
\end{equation}
\end{widetext}
It is noticeable that both $\tilde{U}_{2x}(T)$ and $\tilde{U}_{2z}(T)$ possess a similar mathematical structure to $\tilde{U}_{2y}(T)$ expect their distinct coefficients. If combining them together, we obtain the first order correction to the time evolution operator $\tilde{U}$. Therefore, it turns out that the same forms of cyclic initial state and dynamical phase as those in \Eref{Eq tilde p y'} and \Eref{Eq gamma_pt_2y} remains unchanged. Substituting ${k_y}$ in \Eref{Eq Ly} and \Eref{Eq gamma_pt_2y} with $k={k_y}+{k_x}+{k_z} $  results in the perturbed result. Moreover, they are more accurate results for the AA phase in comparison with the unperturbed and numerical results which is shown in \Fref{Fig AA phase}. Meanwhile, we can see clearly from \Eref{Eq ky},\Eref{Eq kx} and \Eref{Eq kz} that $ \tilde{H}_{2y} $ and $ \tilde{H}_{2x} $ contributes to the higher-order harmonic resonance for $\tilde{\Omega}\approx \omega$ and $\tilde{\Omega}\approx 3\omega$, while $ \tilde{H}_{2z} $ contributes to the higher-order harmonic resonance for $\tilde{\Omega}\approx 2\omega$. 

To evaluate the effectiveness of perturbation theory based on the CHRW method, we depict the AA phase $ \gamma_+ $ and the modulated effective Rabi frequency $ \tilde{\Omega} $ as a function of $\Delta/\omega$ in \Fref{Fig AA phase}, obtained by the numerical method, the CHRW method, and perturbation theory based on the CHRW method. In \Fref{Fig AAe05} and \Fref{Fig AAe15}, we remove the limitation of AA phase to $ [0,2\pi] $ to better exhibit the harmonic resonance peaks of AA phase. It is obvious to see that the results obtained by the CHRW method exhibit an asymptotic tendency for both the AA phase and the modulated effective Rabi frequency. In contrast, the combination of perturbation theory and the CHRW method (denoted as CHRW+PT) accurately depicts the harmonic resonance peaks of $ \gamma_+ $ and sudden changes of $ \tilde{\Omega} $ in the second and the third harmonic resonance regimes, defined as the full width at half maximum (FWHM) of AA phase\cite{RN102}. In \Fref{Fig AAe05} and \Fref{Fig AAe15}, we observe not only the third harmonic resonance driven by the external field, but also the second harmonic resonance arising from the combination of bias and driving field. It is known that the odd higher-order harmonic resonance happens in the symmetric Rabi model. However, the asymmetry in the asymmetric Rabi model induces even higher-order harmonic resonance which is a significant character in the asymmetric Rabi model, compared with the results of the symmetric Rabi model\cite{RN102}. In \Fref{Fig AAe05} where $\epsilon/\omega=0.5$, the second harmonic resonance occurs at $\Delta/\omega \approx 1.75$, and the third harmonic resonance occurs at $\Delta/\omega \approx 2.865$. As the bias increases to $\epsilon/\omega=1.5$ in \Fref{Fig AAe15}, the positions of the second and the third harmonic resonances shift to $\Delta/\omega \approx 1.25$ and $\Delta/\omega \approx 2.515$ respectively, which reflects the compensating effect between $\Delta$ and $\epsilon$ as demonstrated in \Sref{Sec Positions of the harmonic resonance}. Moreover, as shown in Fig. \ref{Fig Omega_e05} and Fig. \ref{Fig Omega_e15}, the second and the third harmonic resonances happen when $ \tilde{\Omega}/\omega \approx 1$ and $ \tilde{\Omega}/\omega \approx 2$ respectively, in agreement with the prediction of the CHRW method.

\section{Hidden symmetry}\label{Sec Hidden symmetry}
Now we come back to discuss the hidden symmetry of the asymmetric semiclassical Rabi model that appears as $ \epsilon=m\omega, m\in \mathbb{Z}$. After careful calculation, we confirm that as $ \epsilon=m\omega, m\in \mathbb{Z}$, the harmonic resonance of geometric quantities is absent for the numerical results at the third harmonic resonance $ \Delta_{\rm{res}} $. Furthermore, we find that in this case, the geometric quantities are actually determined by the arbitrary choices of cyclic initial states. In this section, we first illustrate this discovery using Floquet theory (about which we give a short overview in Appendix \ref{App Floquet theory}). We also explain this phenomenon in terms of hidden symmetry by comparing the asymmetric semiclassical Rabi model with the asymmetric quantum Rabi model.

First of all, one can numerically verify the fact that when $ \epsilon=\omega$ and $\Delta= \Delta_{\rm{res}} $, the quasi-energies satisfy:
\begin{equation}
	q_- - q_+ = m\omega, m\in \mathbb{Z},
\end{equation}
which is equivalent to
\begin{equation} \label{Eq equivalent total phase}
	\theta_+ - \theta_-= 2m\pi, m\in\mathbb{Z}.
\end{equation}
According to Floquet theory, both quasi-energies are physically identical, which means the quasi-energies are degenerate, and any superposition of the cyclic initial states $|\Psi\rangle = c_1 |\psi_{-}(0)\rangle + c_2 |\psi_{+}(0)\rangle $ is also a cyclic initial state with the equivalent quasienergy\cite{RN168}, i.e.,
\begin{equation}
	U(T)|\Psi\rangle=e^{-iq_+T}|\Psi\rangle.
\end{equation}
Because $ |\psi_{-}(0)\rangle $ and $ |\psi_{+}(0)\rangle $ are linear independent, $ |\Psi\rangle $ can be any state vector in the Hilbert space, and therefore the invariant space of $ U(T) $ is the whole space. Thus, we can naturally infer that $ U(T)=e^{-iq_+T}I $. Furthermore, combining \Eref{Eq equivalent total phase} with the complementary relation $\theta_+ + \theta_-= 2l\pi,l\in \mathbb{Z} $, we can set $ \theta_+=\theta_-=0 $, without loss of generality, and in this case, $ U(T)=I $.

Since the cyclic initial state can be, in principle, any vector in the Hilbert space, and geometric quantities which depend on it cannot be uniquely defined, including AA phase and time-energy uncertainty as our focuses in this work. In this sense, the absent resonance obtained numerically is only one of the infinity solutions. It can be verified that the degeneracy of quasi-energies as $\Delta= \Delta_{\rm{res}} $ happens not only when $ \epsilon=\omega $, but when $ \epsilon=m\omega, m\in \mathbb{Z}$ as well.

This intriguing phenomenon stems from the hidden symmetry of the asymmetric Rabi model, which has been reported by a number of researches for the asymmetric quantum Rabi model\cite{RN185,RN256,RN196,RN193}. The Hamiltonian for the asymmetric quantum Rabi model can be written as:
\begin{equation}\label{Eq Hq}
	H_q=\omega a^\dagger a- \frac{g}{4} \sigma_z (a^\dagger +a)-\frac{\Delta}{2}\sigma_x-\frac{\epsilon}{2}\sigma_z,
\end{equation}
where $ \omega $ is the frequency of a quantized resonator field with annihilation and creation operators $ a $ and $ a^\dagger $, and $ g $ is the coupling strength between the qubit and the field. For this Hamiltonian, the hidden symmetry appears as long as $ \epsilon=m\omega, m\in \mathbb{Z}$, and conserved quantity which commutes with the Hamiltonian, and corresponding symmetric unitary operator can be found in the case \cite{RN256,RN196,RN193}.

It is necessary to mention that the coefficients and signs chosen in \Eref{Eq Hq} are corresponding to those in \Eref{Eq classical Rabi H}, and the properties of the asymmetric quantum Rabi model are not reliant on those particular choices. The quantum field in \Eref{Eq Hq} can be reduced to the classical field in \Eref{Eq classical Rabi H} with $ g\left|\langle a \rangle\right|=A$ fixed and $ g\rightarrow 0, \left|\langle a \rangle\right| \rightarrow \infty$, where $\left|\langle a \rangle\right|$ is expectation value of the annihilation operator $ a $ \cite{RN197}.

According to Floquet theory, the Floquet Hamiltonian for the asymmetric semiclassical Rabi model, is
\begin{equation}\label{Eq HF}
	H_F = -\frac{\Delta}{2} \sigma_x - \frac{\epsilon+A\cos(\omega t)}{2} \sigma_z -i\partial_t.
\end{equation}
A comparison of \Eref{Eq HF} and \Eref{Eq Hq} in the matrix form demonstrates the similarity between the asymmetric semiclassical Rabi model and the asymmetric quantum Rabi model. For the asymmetric semiclassical Rabi model, we introduce a set of basis:
\begin{equation}\label{Eq basis for asymmetric semiclassical Rabi model}
	\left| \uparrow \rm{or} \downarrow, n\right\rangle = \left| \uparrow \rm{or} \downarrow \right\rangle e^{-in\omega t},
\end{equation}
where $ \left| \uparrow \right\rangle$ and $ \left| \downarrow \right\rangle $ are the eigenstates of $ \sigma_z $. In the extended Hilbert space $ \mathcal{H} \otimes \mathcal{T}$, those vectors become $ \left| \uparrow \rm{or} \downarrow, n\rangle \right\rangle $ (see Appendix \ref{App Floquet theory} for the details). In this basis, the matrix of $ H_F $ reads:
\begin{widetext}
\begin{equation}\label{Eq HF matrix}
	H_F=\left[
	\begin{array}{c|cccccc}
		\ddots & |\uparrow,-n\rangle\rangle & |\downarrow,-n\rangle\rangle & |\uparrow,-(n+1)\rangle\rangle & |\downarrow,-(n+1)\rangle\rangle & |\uparrow,-(n+2)\rangle\rangle & |\downarrow,-(n+2)\rangle\rangle \\
		\hline|\uparrow,-n\rangle\rangle & n \omega-\frac{\epsilon}{2} & -\frac{\Delta}{2} & -\frac{A}{4} & 0 & 0 & 0 \\
		|\downarrow,-n\rangle\rangle & -\frac{\Delta}{2} & n \omega+\frac{\epsilon}{2} & 0 & \frac{A}{4} & 0 & 0 \\
		|\uparrow,-(n+1)\rangle\rangle & -\frac{A}{4} & 0 & (n+1) \omega-\frac{\epsilon}{2} & -\frac{\Delta}{2} & -\frac{A}{4} & 0 \\
		|\downarrow,-(n+1)\rangle\rangle & 0 & \frac{A}{4} & -\frac{\Delta}{2} & (n+1) \omega+\frac{\epsilon}{2} & 0 & \frac{A}{4} \\
		|\uparrow,-(n+2)\rangle\rangle & 0 & 0 & -\frac{A}{4} & 0 & (n+2) \omega-\frac{\epsilon}{2} & -\frac{\Delta}{2} \\
		|\downarrow,-(n+2)\rangle\rangle & 0 & 0 & 0 & \frac{A}{4} & -\frac{\Delta}{2} & (n+2) \omega+\frac{\epsilon}{2} 
	\end{array}
\right].
\end{equation}

For the asymmetric quantum Rabi model, we introduce another similar set of basis 
\begin{equation}\label{Eq basis for asymmetric quantum Rabi model}
	\left| \uparrow \rm{or} \downarrow, n \right\rangle,
\end{equation}
where  $ \left| \uparrow \right\rangle$ and $ \left| \downarrow \right\rangle $ are the same as the symbols in \Eref{Eq basis for asymmetric semiclassical Rabi model} and $| n \rangle$ is the fork state for the periodic driving field. In this basis, $ H_q$ takes the form
\begin{equation}\label{Eq Hq matrix}
		H_q=\left[
		\begin{array}{c|cccccc}
			\ddots & |\uparrow,n\rangle & |\downarrow,n\rangle & |\uparrow,n+1\rangle & |\downarrow,n+1\rangle & |\uparrow,n+2\rangle & |\downarrow,n+2\rangle \\
			\hline|\uparrow,n\rangle & n \omega-\frac{\epsilon}{2} & -\frac{\Delta}{2} & -\frac{g}{4}\sqrt{n+1} & 0 & 0 & 0 \\
			|\downarrow,n\rangle & -\frac{\Delta}{2} & n \omega+\frac{\epsilon}{2} & 0 & \frac{g}{4}\sqrt{n+1} & 0 & 0 \\
			|\uparrow,n+1\rangle & -\frac{g}{4}\sqrt{n+1} & 0 & (n+1) \omega-\frac{\epsilon}{2} & -\frac{\Delta}{2} & -\frac{g}{4}\sqrt{n+2} & 0 \\
			|\downarrow,n+1\rangle & 0 & \frac{g}{4}\sqrt{n+1} & -\frac{\Delta}{2} & (n+1) \omega+\frac{\epsilon}{2} & 0 & \frac{g}{4}\sqrt{n+2} \\
			|\uparrow,n+2\rangle & 0 & 0 & -\frac{g}{4}\sqrt{n+2} & 0 & (n+2) \omega-\frac{\epsilon}{2} & -\frac{\Delta}{2} \\
			|\downarrow,n+2\rangle & 0 & 0 & 0 & \frac{g}{4}\sqrt{n+2} & -\frac{\Delta}{2} & (n+2) \omega+\frac{\epsilon}{2} 
		\end{array}
		\right].
\end{equation}
\end{widetext}
The structural similarity between $H_F$ and $H_q$ is apparent in the matrix representation: the exactly equal tridiagonal elements for both Hamiltonian (note that $ n\in \mathbb{Z}^{+} $ for both matrices and $-n$ is used in \Eref{Eq HF matrix}).  Meanwhile, the off-tridiagonal elements of $H_q$ depend on $ n $ and those of $ H_F $ depend on $ A $, which is related to the average number of photons in the semiclassical limit\cite{RN197,RN253}. However, in $H_F$, $n$ runs from negative infinity to positive infinity, while in $H_q$, $n$ cannot be negative because it labels the number of photons \cite{RN149}.

\begin{figure*}[htbp]
	\centering
	\subfigure{
		\includegraphics[width=7.5cm]{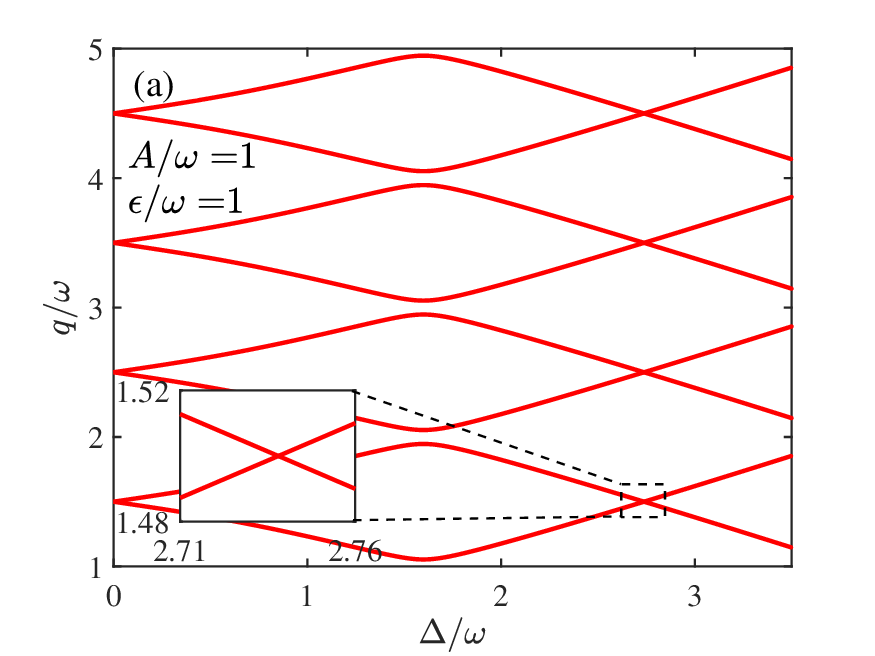}
		\label{Fig quasi_e1}
	}
	\subfigure{
		\includegraphics[width=7.5cm]{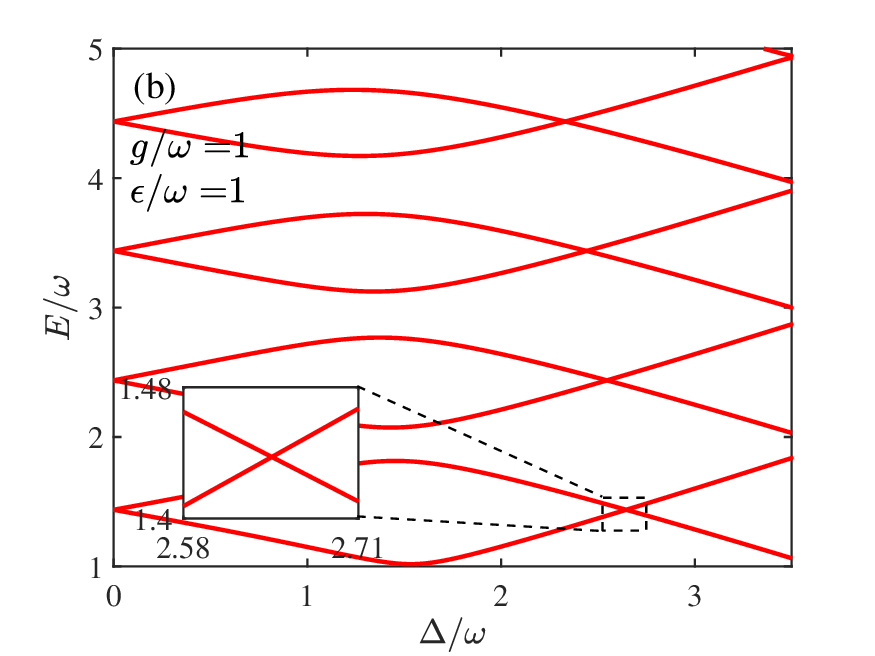}
		\label{Fig eign_e1}
	}
	\subfigure{
		\includegraphics[width=7.5cm]{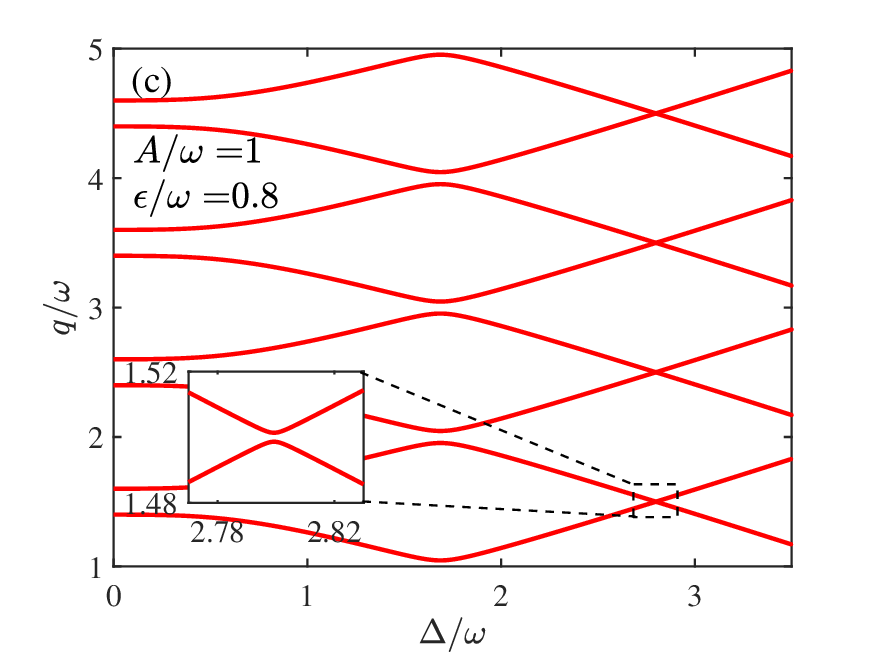}
		\label{Fig quasi_e08}
	}
	\subfigure{
		\includegraphics[width=7.5cm]{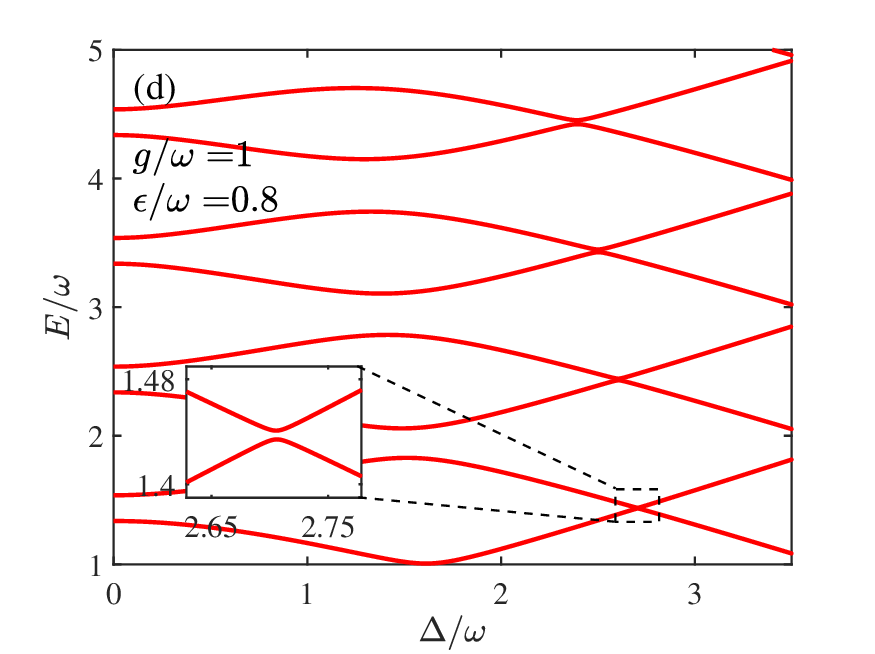}
		\label{Fig eigen_e08}
	}
	\caption{Comparison of (quasi-)energy spectrum as a function of $ \Delta/\omega $ between the asymmetric semiclassical Rabi model and the asymmetric quantum Rabi model under different parameters. (a) and (c) Quasi-energy spectrum of the asymmetric semiclassical Rabi model with $ A/\omega=1, \epsilon/\omega=1,0.8 $. (b) and (d) Eigen-energy spectrum of the asymmetric quantum Rabi model with $ g/\omega=1, \epsilon/\omega=1,0.8 $. For $ \epsilon/\omega=1 $, the quasi-energies or eigen-energies are degenerate at certain points, while no degeneracy occurs for $ \epsilon/\omega=0.8$. 
	}
	\label{Fig comparison}
\end{figure*}

With the matrices of both Hamiltonian, we calculate the convergence values of the eigenvalues to obtain accurate energy spectra by incrementally increasing their dimensions. Figure \ref{Fig quasi_e1} and Figure \ref{Fig quasi_e08} display the quasienergy of the asymmetric semiclassical Rabi model against $ \Delta/\omega $ with $ A/\omega=1$ for $\epsilon/\omega=1 $ and $\epsilon/\omega=0.8 $, respectively, while Figure \ref{Fig eign_e1} and Figure \ref{Fig eigen_e08} the eigenenergy of the asymmetric quantum Rabi model with $ g/\omega=1$, for $\epsilon/\omega=1 $ and $\epsilon/\omega=0.8$. From these figures, we can see that for $ \epsilon/\omega=1 $, the quasi-energies are degenerate at $  \Delta_{\rm{res}}/\omega\approx 2.74 $, and degeneracies can also be observed for the eigenenergy spectrum in the vicinity of $  \Delta/\omega\approx 2.5 $. However, no level crossing occurs when $ \epsilon/\omega\ne 1$. Though not shown here, we also verified the general existence of level crossings for $ \epsilon/\omega= m,m\in \mathbb{Z}$, and anti-crossings for $ \epsilon/\omega \ne m$, and we found that these characters were not influenced by the value of $ A/\omega $ or $ g/\omega $, which mainly shifted the spectra along the $ \Delta/\omega $ axis. This fact provides strong evidence for the hidden symmetry of both the asymmetric semiclassical Rabi model and the asymmetric quantum Rabi model as structural properties of the Hamiltonian, which probably have impacts in all coupling regimes of light-matter interaction or driven two-level system. In consequence, the hidden symmetry contributes to the non-uniqueness of the geometric quantities, AA phase and time-energy uncertainty.

\section{Conclusion}\label{Sec Conclusion}
In this work, we investigate the harmonic resonance of AA phase and time-energy uncertainty as geometric quantities and hidden symmetry in the asymmetric Rabi model by numerical method. At the same time, we apply perturbation theory based on the CHRW method to analytically give accurate time evolution operator and then calculate accurate geometric phase which is in good agreement with numerically exact results.  The asymmetry in the Rabi model have important effects on the geometric phase. In comparison with the existence of only odd harmonic resonances at the absence of bias,  various resonance phenomena take place at the presence of bias, including the emergence of the even higher-order harmonic resonances induced by the asymmetry, and the disappearance of odd higher-order resonance caused by the hidden symmetry present when $\epsilon=m\omega,m\in \mathbb{Z}$. Besides,
with the increase of the bias, all the harmonic resonance shift toward smaller values of $ \Delta/\omega $ owing to the compensating effect between $\epsilon$ and $\Delta$, confirmed by both the numerical method and the CHRW method using the condition $\tilde{\Omega}=m\omega, m\in \mathbb{Z}$. 

We illustrate the features of harmonic resonance encapsulated by the repeated intersection of AA phase at $\gamma=\pi$, attainment of a local maximum in time-energy uncertainty, and rapid oscillations in the population of the spin-up state, $\rm{P_{up}}$.  The harmonic resonance was mainly steered by the initial state $|\psi (0)\rangle$, which is highly sensitive to $\Delta/\omega$ in the vicinity of harmonic resonance. The results obtained by the CHRW method, in which the influences of counterrotating terms and bias were all taken into account in the rotating frame, exhibit asymptotic tendencies for both AA phase and modulated effective Rabi frequency. With the combination of perturbation theory, we further effectively depicting the harmonic resonance peaks of $ \gamma_+ $ and sudden changes of $ \tilde{\Omega} $ in the second and the third harmonic resonance regimes. 

Upon investigation of the absence of harmonic resonance in the results of numerical calculation for $\epsilon=m\omega,m \in \mathbb{Z},\Delta= \Delta_{\rm{res}} $, we shed light on the dependence of geometric quantities on the choice of cyclic initial states due to the degeneracy of quasienergy of the asymmetric semiclassical Rabi model. Using Floquet theory, we found the structural similarity between $H_F$ and $H_q$ in the matrix representation and spectra, which reveals the hidden symmetry in the asymmetric semiclassical Rabi model as an analogy of that in the asymmetric quantum Rabi model reported already in \cite{RN185,RN256,RN196,RN193}. Our results suggest that the hidden symmetry reflects the intrinsic property of asymmetric Rabi model regardless of the amplitude of driving field or the coupling strength between the qubit and oscillating field, which plays a potential role in all coupling regimes of light-matter interaction or driven two-level system.  Meanwhile, since the the time-energy uncertainty equals the length of path traversed by the state on the Bloch sphere, it can be measured in superconducting qubit systems via quantum process tomography.

\begin{acknowledgments}
	The work is supported by National Natural Science Foundation of China (Grants No. 11774226 and No. 61927822).
\end{acknowledgments}

	\appendix
\section{Projective Hilbert space }\label{App Projective Hilbert space}
In quantum mechanics, states are described by vectors in the Hilbert space $ \mathcal{H} $. However, any two vectors $ \psi,\phi \in \mathcal{H}  $, satisfying
\begin{equation}
	\psi=c\phi,c\in \mathbb{C},
\end{equation}
are physically equivalent $ (\psi\sim\phi) $. Therefore, the projective Hilbert space $ \mathcal{P} $, defined as:
\begin{equation}
	\mathcal{P} :=\mathcal{H}/\sim
\end{equation}
is the proper space for representing quantum states that can be physically distinguished.

\section{Analysis of hidden symmetry}\label{App analysis of HS}
\subsection{Floquet theory}\label{App Floquet theory}
For a time-dependent Hamiltonian which satisfies $ H(t+T)=H(t) $, the solution to the Schrödinger equation 
\begin{equation}\label{Schrödinger equation}
	i\partial_t |\psi(t)\rangle = H(t) |\psi(t)\rangle
\end{equation}
takes the form:
\begin{equation}\label{Eq floquet}
	|\psi_\alpha(t)\rangle= |u_\alpha(t)\rangle e^{-iq_\alpha t},
\end{equation}
where $ q_\alpha $ is called quasienergy, and $ |u_\alpha(t+T)\rangle = |u_\alpha(t)\rangle $. This is often referred to as Floquet theorem. Since $ |\psi_\alpha(0)\rangle $ merely acquires a phase factor after $ t=T $, i.e., the total phase $ \theta = -q_\alpha T $, it is called a cyclic initial state. Inserting \Eref{Eq floquet} into \Eref{Schrödinger equation}, the original problem can be transformed into an eigenvalue problem with Floquet Hamiltonian $ H_F $:
\begin{equation}
	H_F \left|u_\alpha(t)\right\rangle =\left[H(t)-i\partial_t\right]|u_\alpha(t)\rangle = q_\alpha |u_\alpha(t)\rangle.
\end{equation}
Note that the subscript $ \alpha $ is to label the order of the eigenvectors or eigenvalues, but there are actually infinite set of solutions, because $ \left|u_{\alpha}(t),n\right\rangle = e^{-in\omega t} \left|u_{\alpha}(t),0\right\rangle \equiv e^{-in\omega t} \left|u_{\alpha}(t)\right\rangle $ is physically identical to \Eref{Eq floquet} only with the shifted quasienergy $ q_{\alpha,n}\equiv q_\alpha-n\omega $. Thus, it is sufficient to consider the set of quasi-energies within an interval of width $ \omega $ and corresponding total phases within an interval of width $ 2\pi $.

More importantly, $ H_F $ can be written as a time-independent infinite matrix through the expansion of the original Hilbert space $ \mathcal{H} $ to $ \mathcal{H} \otimes \mathcal{T}$\cite{RN149}, where $ \mathcal{T} $ contains all T-periodic functions. A particular set of orthonormal complete basis in $ \mathcal{T} $ is written as $ \{|n\rangle\} $ with the plane wave function
\begin{equation}
	\langle t|n \rangle = e^{-in\omega t}.
\end{equation}

Considering a T-periodic state vector $ \left|u_\alpha(t),n\right\rangle \in \mathcal{H} $, we can expand it in a Fourier series, plane wave terms are derived:
\begin{equation}
	\left|u_{\alpha}(t), n\right\rangle=e^{-i n \omega t}\left|u_\alpha(t)\right\rangle=\sum_l e^{-i l \omega t}|u_\alpha^{(n-l)}\rangle,
\end{equation}
where $ |u_\alpha^{(k)}\rangle $ are time-independent Fourier coefficients. Now in $ \mathcal{H} \otimes \mathcal{T}$, the corresponding state vector is defined as
\begin{equation}
	\left| u_\alpha, n\rangle \right\rangle \equiv \sum_l\left|u_\alpha^{(n-l)}\right\rangle \otimes \left|l\right\rangle,
\end{equation}
and finally the inner product is defined as
\begin{equation}
	\langle \langle u_{\alpha},n|u_{\beta},m\rangle\rangle \equiv \frac{1}{T}\int_{0}^{T} dt \left\langle u_{\alpha}(t),n|u_{\beta}(t),m\right\rangle.
\end{equation}

\subsection{Complementary relation of AA phases}\label{App Complementary}
To derive the complementary relation of AA phases, we start from the Schr\"odinger equation in \Eref{Eq SchrodingerU}. For this purpose, here we denote the Hamiltonian as $ H(t)=[h_1,h_2;h_2,-h_1]$, where $ h_1=-\left(A \cos \omega t+\epsilon\right)/2, h_2=-\Delta/2$, and $ U(t)=[u_1,u_2;u_3,u_4] $.
Then we can list the following ordinary differential equations (ODE's):
\begin{eqnarray}
	i\partial_t u_1=h_1u_1+h_2u_3,  \label{Eq a}\\
	i\partial_t u_2=h_1u_2+h_2u_4,  \label{Eq b}\\
	i\partial_t u_3=h_2u_1-h_1u_3,  \label{Eq c}\\
	i\partial_t u_4=h_2u_2-h_1u_4,  \label{Eq d}
\end{eqnarray}
with boundary conditions:
\begin{eqnarray}
	u_1(0)=u_4(0)=1, \nonumber\\
	u_2(0)=u_3(0)=0. \nonumber
\end{eqnarray}
Note that, if we substitute the variables $ u_1 $ and $ u_2 $ in \Eref{Eq c} and \Eref{Eq d} with $ u_4^* $ and $ -u_3^* $ respectively and take the conjugate of both equations, we will have the same equations as \Eref{Eq a} and \Eref{Eq b}, with the boundary conditions still satisfied. Due to the uniqueness theorem for ODE's, the relations
\begin{equation}\label{Eq elementU}
	u_1=  u_4^*, \quad  u_2= -u_3^*
\end{equation}
must be kept in the solution for \Eref{Eq SchrodingerU}. Since the evolution operator is unitary, we can get
\begin{equation}
\begin{split}
	U(t)U(t)^{\dagger}&= 
\left[\begin{array}{cc}
	u_1 & u_2 \\
	-u_2^* & u_1^*
\end{array}\right] 
\left[\begin{array}{cc}
	u_1^* & -u_2 \\
	u_2^* & u_1
\end{array}\right]\\
&=\left[\begin{array}{cc}
	\left|u_1\right|^2 +\left|u_2\right|^2 & 0 \\
	0 & \left|u_1\right|^2 +\left|u_2\right|^2
\end{array}\right]\\
&=I,
\end{split}
\end{equation}
which implies that
\begin{equation}\label{Eq detU}
	det\left[U(t)\right] =\left|u_1\right|^2 +\left|u_2\right|^2=1, \forall t\ge 0.
\end{equation}
Because $ e^{i\theta_+} $ and $ e^{i\theta-} $ are the two eigenvalues of $ U(T) $, $e^{i\theta_+} e^{i\theta-} = e^{i \theta_+ + \theta_-}=det\left[U(T)\right]=1$. Therefore, we get the relation between total phases
\begin{equation}\label{Eq complementary relation for total phase}
	\theta_+ + \theta_-= 2l\pi,l\in \mathbb{Z}.
\end{equation}

Moreover, with \Eref{Eq elementU} and \Eref{Eq detU} it is easy to prove through some calculations that the two eigenvectors for $ U(T) $, i.e., the cyclic initial states $ |\psi_{\pm}(0)\rangle $ are orthogonal to each other, and thus
\begin{equation}
	\langle \psi_{-}(0)|U^{\dagger}HU|\psi_{-}(0)\rangle=-\langle \psi_{+}(0)|U^{\dagger}HU|\psi_{+}(0)\rangle,
\end{equation}
which leads to 
\begin{equation}\label{Eq complementary relation for dynamical phase}
	\alpha_-=-\alpha_+.
\end{equation}
Finally we obtain from \Eref{Eq definition for AA phase},\Eref{Eq definition for dynamical phase},\Eref{Eq complementary relation for total phase} and \Eref{Eq complementary relation for dynamical phase} that $ \gamma_+ + \gamma_- = 2l\pi,l\in \mathbb{Z} $. 

\section{Perturbation theory}\label{App pt}
Near the third harmonic resonance, $ \tilde{\Omega}\approx 2\omega $. Thus, the total phase given in \Eref{Eq total_phase pt} can be simplified as 
\begin{equation}
	\theta_+'\approx  \frac{\sqrt{1+k^2}\left(\tilde{\Omega}-2\omega\right)T-\omega T}{2}.
\end{equation}
By $ \theta_{+}'= \frac{\tilde{\Omega}'}{2}T+\frac{2n+1}{2}\omega T, n\in \mathbb{Z} $, we get the modified Rabi frequency
\begin{equation}
	\tilde{\Omega}'\approx \sqrt{1+k^2}\left(\tilde{\Omega}-2\omega\right)+2\omega.
\end{equation}
To calculate the dynamical phases, our target is to solve the integral in \Eref{Eq alpha formula}, since the total phases given by the CHRW method are reserved due to their tiny variation. In the first step, we write the middle part of the integrand
\begin{widetext}
\begin{equation}\label{Eq DeHeD}
	\begin{split}
		&D^{\dagger}e^SHe^{-S}D=-\frac{\tau_y}{2} \Bigg\{ \frac{ \sin\Theta}{x}\left[ \nu(u\varepsilon-v\Delta)-\mu(u\Delta +v\varepsilon) \right] \Bigg\} -\frac{\tau_x}{2}\Bigg\{ \cos^2\left(\frac{\Theta}{2}\right)\left( v^2\Delta-2uv\varepsilon-u^2\Delta \right)+\frac{1}{x^2}\sin^2\left(\frac{\Theta}{2}\right)\\
		&\left( \Delta\left( \mu^2-\nu^2 \right) -2\varepsilon  \mu\nu \right)\Bigg\}-\frac{\tau_z}{2}\Bigg\{ \cos^2\left(\frac{\Theta}{2}\right)(u^2 \varepsilon-2uv\Delta-v^2\varepsilon)+\frac{1}{x^2}\sin^2\left(\frac{\Theta}{2}\right)\left[ \varepsilon\left( \mu^2-\nu^2 \right)+2\Delta  \mu \nu\right] \Bigg\}.
	\end{split}
\end{equation}

Next, we calculate the state vector at time $t$, using the evolution operator $ \tilde{U} $ without perturbation

\begin{equation}\label{Eq U_tilde p_tilde'}
	\tilde{U} |\tilde{\psi}_{+}'(0) \rangle = \frac{1}{{L}}
	\left(\begin{array}{c}
		e^{-i\omega t/2}\left[ (\tilde{A}+{k}\tilde{\delta})\cos \left(\frac{\tilde{\Omega}t}{2}\right)+i(\tilde{A}\sqrt{1+{k}^2}-{k}\tilde{\Omega})\sin \left(\frac{\tilde{\Omega}t}{2}\right) \right] \\
		e^{i\omega t/2}\left[ ({k}\tilde{A}-\tilde{\delta}-\sqrt{1+{k}^2}\tilde{\Omega})\cos \left(\frac{\tilde{\Omega}t}{2}\right)-i(\tilde{\Omega}+\sqrt{1+{k}^2}\tilde{\delta})\sin \left(\frac{\tilde{\Omega}t}{2}\right) \right]
	\end{array}\right),
\end{equation}
where the initial state under perturbation $ |\tilde{+}' \rangle $ takes the form of \Eref{Eq tilde p y'} with all components $ \tau_x,\tau_y $ and $ \tau_z $ taken into consideration.

To calculate the integral in \Eref{Eq alpha formula} analytically, we also use $ \omega $ to approximate $ \tilde{\Omega} $, which leads to negligible effects. Last but not least, in the process, Taylor expansion is adopted
\begin{equation}
	\cos\left(\frac{\Theta}{2}\right)\approx 1-\frac{1}{2}\left(\frac{\Theta}{2}\right)^2,\sin \left(\frac{\Theta}{2}\right)\approx \frac{\Theta}{2}.
\end{equation}
Combining \Eref{Eq DeHeD} and \Eref{Eq U_tilde p_tilde'} with the approximations above, we then calculate the dynamical phase given by perturbation theory
\begin{equation}\label{Eq alpha_p21}
\begin{split}
&\alpha_+'
\approx- \int_{0}^{T} \langle \tilde{\psi}_{+}'(0)|\tilde{U}^{\dagger}D^{\dagger}e^SHe^{-S}D\tilde{U}|\tilde{\psi}_{+}'(0) \rangle d\tau \approx \frac{\pi}{2{L}^2} \Bigg\{ (\tilde{\delta}+\sqrt{{k}^2+1}\tilde{\Omega}-{k}\tilde{A}) \bigg\{\frac{A^2 \tilde{\delta}}{2\omega^3}\left[\epsilon(\nu^2-\mu^2)-2\Delta \mu \nu\right]\\
&-\frac{\tilde{\delta}}{128\omega}(3z^4-64z^2+512)(u^2\epsilon-2uv\Delta-v^2\epsilon)-\frac{2A\tilde{A}}{\omega^2}\left[(u\epsilon-v\Delta)\nu-(u\Delta+v\epsilon)\mu\right]+\frac{A\tilde{A}}{\omega}\left[\frac{2}{z}J_1(z)+1\right][k(u^2-v^2)+2uv]\bigg\}	\\
&+\frac{4{k}}{x^2}J_2(z)\left(u\nu-v\mu\right)\left[{k}\tilde{A}(\tilde{\Omega}+\tilde{\delta})-(1+\sqrt{{k}^2+1})\tilde{\Omega}\tilde{\delta}-\sqrt{{k}^2+1}\tilde{\Omega}^2-\tilde{\delta}^2\right]\Bigg\}.
\end{split}
\end{equation}
\end{widetext}

\bibliography{Geo}

\begin{thebibliography}{68}%
\makeatletter
\providecommand \@ifxundefined [1]{%
 \@ifx{#1\undefined}
}%
\providecommand \@ifnum [1]{%
 \ifnum #1\expandafter \@firstoftwo
 \else \expandafter \@secondoftwo
 \fi
}%
\providecommand \@ifx [1]{%
 \ifx #1\expandafter \@firstoftwo
 \else \expandafter \@secondoftwo
 \fi
}%
\providecommand \natexlab [1]{#1}%
\providecommand \enquote  [1]{``#1''}%
\providecommand \bibnamefont  [1]{#1}%
\providecommand \bibfnamefont [1]{#1}%
\providecommand \citenamefont [1]{#1}%
\providecommand \href@noop [0]{\@secondoftwo}%
\providecommand \href [0]{\begingroup \@sanitize@url \@href}%
\providecommand \@href[1]{\@@startlink{#1}\@@href}%
\providecommand \@@href[1]{\endgroup#1\@@endlink}%
\providecommand \@sanitize@url [0]{\catcode `\\12\catcode `\$12\catcode
  `\&12\catcode `\#12\catcode `\^12\catcode `\_12\catcode `\%12\relax}%
\providecommand \@@startlink[1]{}%
\providecommand \@@endlink[0]{}%
\providecommand \url  [0]{\begingroup\@sanitize@url \@url }%
\providecommand \@url [1]{\endgroup\@href {#1}{\urlprefix }}%
\providecommand \urlprefix  [0]{URL }%
\providecommand \Eprint [0]{\href }%
\providecommand \doibase [0]{https://doi.org/}%
\providecommand \selectlanguage [0]{\@gobble}%
\providecommand \bibinfo  [0]{\@secondoftwo}%
\providecommand \bibfield  [0]{\@secondoftwo}%
\providecommand \translation [1]{[#1]}%
\providecommand \BibitemOpen [0]{}%
\providecommand \bibitemStop [0]{}%
\providecommand \bibitemNoStop [0]{.\EOS\space}%
\providecommand \EOS [0]{\spacefactor3000\relax}%
\providecommand \BibitemShut  [1]{\csname bibitem#1\endcsname}%
\let\auto@bib@innerbib\@empty
%</preamble>
\bibitem [{\citenamefont {Berry}(1984)}]{RN129}%
  \BibitemOpen
  \bibfield  {author} {\bibinfo {author} {\bibfnamefont {M.~V.}\ \bibnamefont
  {Berry}},\ }\href {https://doi.org/DOI 10.1098/rspa.1984.0023} {\bibfield
  {journal} {\bibinfo  {journal} {Proceedings of the Royal Society of London.
  A. Mathematical and Physical Sciences}\ }\textbf {\bibinfo {volume} {392}},\
  \bibinfo {pages} {45} (\bibinfo {year} {1984})}\BibitemShut {NoStop}%
\bibitem [{\citenamefont {Leek}\ \emph {et~al.}(2007)\citenamefont {Leek},
  \citenamefont {Fink}, \citenamefont {Blais}, \citenamefont {Bianchetti},
  \citenamefont {Goppl}, \citenamefont {Gambetta}, \citenamefont {Schuster},
  \citenamefont {Frunzio}, \citenamefont {Schoelkopf},\ and\ \citenamefont
  {Wallraff}}]{RN115}%
  \BibitemOpen
  \bibfield  {author} {\bibinfo {author} {\bibfnamefont {P.~J.}\ \bibnamefont
  {Leek}}, \bibinfo {author} {\bibfnamefont {J.~M.}\ \bibnamefont {Fink}},
  \bibinfo {author} {\bibfnamefont {A.}~\bibnamefont {Blais}}, \bibinfo
  {author} {\bibfnamefont {R.}~\bibnamefont {Bianchetti}}, \bibinfo {author}
  {\bibfnamefont {M.}~\bibnamefont {Goppl}}, \bibinfo {author} {\bibfnamefont
  {J.~M.}\ \bibnamefont {Gambetta}}, \bibinfo {author} {\bibfnamefont {D.~I.}\
  \bibnamefont {Schuster}}, \bibinfo {author} {\bibfnamefont {L.}~\bibnamefont
  {Frunzio}}, \bibinfo {author} {\bibfnamefont {R.~J.}\ \bibnamefont
  {Schoelkopf}},\ and\ \bibinfo {author} {\bibfnamefont {A.}~\bibnamefont
  {Wallraff}},\ }\href {https://doi.org/10.1126/science.1149858} {\bibfield
  {journal} {\bibinfo  {journal} {Science}\ }\textbf {\bibinfo {volume}
  {318}},\ \bibinfo {pages} {1889} (\bibinfo {year} {2007})}\BibitemShut
  {NoStop}%
\bibitem [{\citenamefont {Thouless}\ \emph {et~al.}(1982)\citenamefont
  {Thouless}, \citenamefont {Kohmoto}, \citenamefont {Nightingale},\ and\
  \citenamefont {den Nijs}}]{RN150}%
  \BibitemOpen
  \bibfield  {author} {\bibinfo {author} {\bibfnamefont {D.~J.}\ \bibnamefont
  {Thouless}}, \bibinfo {author} {\bibfnamefont {M.}~\bibnamefont {Kohmoto}},
  \bibinfo {author} {\bibfnamefont {M.~P.}\ \bibnamefont {Nightingale}},\ and\
  \bibinfo {author} {\bibfnamefont {M.}~\bibnamefont {den Nijs}},\ }\href
  {https://doi.org/10.1103/PhysRevLett.49.405} {\bibfield  {journal} {\bibinfo
  {journal} {Physical Review Letters}\ }\textbf {\bibinfo {volume} {49}},\
  \bibinfo {pages} {405} (\bibinfo {year} {1982})}\BibitemShut {NoStop}%
\bibitem [{\citenamefont {Oka}\ and\ \citenamefont {Aoki}(2009)}]{RN151}%
  \BibitemOpen
  \bibfield  {author} {\bibinfo {author} {\bibfnamefont {T.}~\bibnamefont
  {Oka}}\ and\ \bibinfo {author} {\bibfnamefont {H.}~\bibnamefont {Aoki}},\
  }\href {https://doi.org/10.1103/PhysRevB.79.081406} {\bibfield  {journal}
  {\bibinfo  {journal} {Physical Review B}\ }\textbf {\bibinfo {volume} {79}},\
  \bibinfo {pages} {081406} (\bibinfo {year} {2009})}\BibitemShut {NoStop}%
\bibitem [{\citenamefont {Cronin}\ \emph {et~al.}(2009)\citenamefont {Cronin},
  \citenamefont {Schmiedmayer},\ and\ \citenamefont {Pritchard}}]{RN152}%
  \BibitemOpen
  \bibfield  {author} {\bibinfo {author} {\bibfnamefont {A.~D.}\ \bibnamefont
  {Cronin}}, \bibinfo {author} {\bibfnamefont {J.}~\bibnamefont
  {Schmiedmayer}},\ and\ \bibinfo {author} {\bibfnamefont {D.~E.}\ \bibnamefont
  {Pritchard}},\ }\href {https://doi.org/10.1103/RevModPhys.81.1051} {\bibfield
   {journal} {\bibinfo  {journal} {Reviews of Modern Physics}\ }\textbf
  {\bibinfo {volume} {81}},\ \bibinfo {pages} {1051} (\bibinfo {year}
  {2009})}\BibitemShut {NoStop}%
\bibitem [{\citenamefont {Aharonov}\ and\ \citenamefont {Bohm}(1959)}]{RN153}%
  \BibitemOpen
  \bibfield  {author} {\bibinfo {author} {\bibfnamefont {Y.}~\bibnamefont
  {Aharonov}}\ and\ \bibinfo {author} {\bibfnamefont {D.}~\bibnamefont
  {Bohm}},\ }\href {https://doi.org/10.1103/PhysRev.115.485} {\bibfield
  {journal} {\bibinfo  {journal} {Physical Review}\ }\textbf {\bibinfo {volume}
  {115}},\ \bibinfo {pages} {485} (\bibinfo {year} {1959})}\BibitemShut
  {NoStop}%
\bibitem [{\citenamefont {Resta}(2000)}]{RN283}%
  \BibitemOpen
  \bibfield  {author} {\bibinfo {author} {\bibfnamefont {R.}~\bibnamefont
  {Resta}},\ }\href {https://doi.org/Doi 10.1088/0953-8984/12/9/201} {\bibfield
   {journal} {\bibinfo  {journal} {Journal of Physics-Condensed Matter}\
  }\textbf {\bibinfo {volume} {12}},\ \bibinfo {pages} {R107} (\bibinfo {year}
  {2000})}\BibitemShut {NoStop}%
\bibitem [{\citenamefont {Xie}\ \emph {et~al.}(2021)\citenamefont {Xie},
  \citenamefont {Pu}, \citenamefont {Jin}, \citenamefont {Xu}, \citenamefont
  {Guo}, \citenamefont {Li}, \citenamefont {Gao}, \citenamefont {Ma},\ and\
  \citenamefont {Luo}}]{RN307}%
  \BibitemOpen
  \bibfield  {author} {\bibinfo {author} {\bibfnamefont {X.}~\bibnamefont
  {Xie}}, \bibinfo {author} {\bibfnamefont {M.}~\bibnamefont {Pu}}, \bibinfo
  {author} {\bibfnamefont {J.}~\bibnamefont {Jin}}, \bibinfo {author}
  {\bibfnamefont {M.}~\bibnamefont {Xu}}, \bibinfo {author} {\bibfnamefont
  {Y.}~\bibnamefont {Guo}}, \bibinfo {author} {\bibfnamefont {X.}~\bibnamefont
  {Li}}, \bibinfo {author} {\bibfnamefont {P.}~\bibnamefont {Gao}}, \bibinfo
  {author} {\bibfnamefont {X.}~\bibnamefont {Ma}},\ and\ \bibinfo {author}
  {\bibfnamefont {X.}~\bibnamefont {Luo}},\ }\href
  {https://doi.org/10.1103/PhysRevLett.126.183902} {\bibfield  {journal}
  {\bibinfo  {journal} {Physical Review Letters}\ }\textbf {\bibinfo {volume}
  {126}},\ \bibinfo {pages} {183902} (\bibinfo {year} {2021})}\BibitemShut
  {NoStop}%
\bibitem [{\citenamefont {Leinonen}\ \emph {et~al.}(2023)\citenamefont
  {Leinonen}, \citenamefont {Hannonen}, \citenamefont {Partanen}, \citenamefont
  {Heikkinen}, \citenamefont {Setälä}, \citenamefont {Friberg},\ and\
  \citenamefont {Hakala}}]{RN362}%
  \BibitemOpen
  \bibfield  {author} {\bibinfo {author} {\bibfnamefont {A.}~\bibnamefont
  {Leinonen}}, \bibinfo {author} {\bibfnamefont {A.}~\bibnamefont {Hannonen}},
  \bibinfo {author} {\bibfnamefont {H.}~\bibnamefont {Partanen}}, \bibinfo
  {author} {\bibfnamefont {J.}~\bibnamefont {Heikkinen}}, \bibinfo {author}
  {\bibfnamefont {T.}~\bibnamefont {Setälä}}, \bibinfo {author}
  {\bibfnamefont {A.~T.}\ \bibnamefont {Friberg}},\ and\ \bibinfo {author}
  {\bibfnamefont {T.~K.}\ \bibnamefont {Hakala}},\ }\href
  {https://doi.org/10.1038/s42005-023-01249-2} {\bibfield  {journal} {\bibinfo
  {journal} {Communications Physics}\ }\textbf {\bibinfo {volume} {6}},\
  \bibinfo {pages} {132} (\bibinfo {year} {2023})}\BibitemShut {NoStop}%
\bibitem [{\citenamefont {Yi}\ and\ \citenamefont {Chang}(2004)}]{add5}%
  \BibitemOpen
  \bibfield  {author} {\bibinfo {author} {\bibfnamefont {X.~X.}\ \bibnamefont
  {Yi}}\ and\ \bibinfo {author} {\bibfnamefont {J.~L.}\ \bibnamefont {Chang}},\
  }\href {https://doi.org/10.1103/PhysRevA.70.012108} {\bibfield  {journal}
  {\bibinfo  {journal} {Physical Review A}\ }\textbf {\bibinfo {volume} {70}},\
  \bibinfo {pages} {012108} (\bibinfo {year} {2004})}\BibitemShut {NoStop}%
\bibitem [{\citenamefont {Ying}\ \emph {et~al.}(2020)\citenamefont {Ying},
  \citenamefont {Gentile}, \citenamefont {Baltan\'as}, \citenamefont
  {Frustaglia}, \citenamefont {Ortix},\ and\ \citenamefont {Cuoco}}]{add6}%
  \BibitemOpen
  \bibfield  {author} {\bibinfo {author} {\bibfnamefont {Z.-J.}\ \bibnamefont
  {Ying}}, \bibinfo {author} {\bibfnamefont {P.}~\bibnamefont {Gentile}},
  \bibinfo {author} {\bibfnamefont {J.~P.}\ \bibnamefont {Baltan\'as}},
  \bibinfo {author} {\bibfnamefont {D.}~\bibnamefont {Frustaglia}}, \bibinfo
  {author} {\bibfnamefont {C.}~\bibnamefont {Ortix}},\ and\ \bibinfo {author}
  {\bibfnamefont {M.}~\bibnamefont {Cuoco}},\ }\href
  {https://doi.org/10.1103/PhysRevResearch.2.023167} {\bibfield  {journal}
  {\bibinfo  {journal} {Phys. Rev. Res.}\ }\textbf {\bibinfo {volume} {2}},\
  \bibinfo {pages} {023167} (\bibinfo {year} {2020})}\BibitemShut {NoStop}%
\bibitem [{\citenamefont {Mead}(1992)}]{RN284}%
  \BibitemOpen
  \bibfield  {author} {\bibinfo {author} {\bibfnamefont {C.~A.}\ \bibnamefont
  {Mead}},\ }\href {https://doi.org/DOI 10.1103/RevModPhys.64.51} {\bibfield
  {journal} {\bibinfo  {journal} {Reviews of Modern Physics}\ }\textbf
  {\bibinfo {volume} {64}},\ \bibinfo {pages} {51} (\bibinfo {year}
  {1992})}\BibitemShut {NoStop}%
\bibitem [{\citenamefont {Aharonov}\ and\ \citenamefont
  {Anandan}(1987)}]{RN128}%
  \BibitemOpen
  \bibfield  {author} {\bibinfo {author} {\bibfnamefont {Y.}~\bibnamefont
  {Aharonov}}\ and\ \bibinfo {author} {\bibfnamefont {J.}~\bibnamefont
  {Anandan}},\ }\href {https://doi.org/10.1103/PhysRevLett.58.1593} {\bibfield
  {journal} {\bibinfo  {journal} {Physical Review Letters}\ }\textbf {\bibinfo
  {volume} {58}},\ \bibinfo {pages} {1593} (\bibinfo {year}
  {1987})}\BibitemShut {NoStop}%
\bibitem [{\citenamefont {Anandan}(1988)}]{RN137}%
  \BibitemOpen
  \bibfield  {author} {\bibinfo {author} {\bibfnamefont {J.}~\bibnamefont
  {Anandan}},\ }\href {<Go to ISI>://WOS:A1988T139700002} {\bibfield  {journal}
  {\bibinfo  {journal} {Annales De L Institut Henri Poincare-Physique
  Theorique}\ }\textbf {\bibinfo {volume} {49}},\ \bibinfo {pages} {271}
  (\bibinfo {year} {1988})}\BibitemShut {NoStop}%
\bibitem [{\citenamefont {Zhao}\ \emph {et~al.}(2021)\citenamefont {Zhao},
  \citenamefont {Dong}, \citenamefont {Zhang}, \citenamefont {Guo},
  \citenamefont {Tong},\ and\ \citenamefont {Yin}}]{RN305}%
  \BibitemOpen
  \bibfield  {author} {\bibinfo {author} {\bibfnamefont {P.~Z.}\ \bibnamefont
  {Zhao}}, \bibinfo {author} {\bibfnamefont {Z.~J.~Z.}\ \bibnamefont {Dong}},
  \bibinfo {author} {\bibfnamefont {Z.~X.}\ \bibnamefont {Zhang}}, \bibinfo
  {author} {\bibfnamefont {G.~P.}\ \bibnamefont {Guo}}, \bibinfo {author}
  {\bibfnamefont {D.~M.}\ \bibnamefont {Tong}},\ and\ \bibinfo {author}
  {\bibfnamefont {Y.}~\bibnamefont {Yin}},\ }\href {https://doi.org/ARTN 250362
  10.1007/s11433-020-1641-1} {\bibfield  {journal} {\bibinfo  {journal}
  {Science China-Physics Mechanics \& Astronomy}\ }\textbf {\bibinfo {volume}
  {64}},\ \bibinfo {pages} {250362} (\bibinfo {year} {2021})}\BibitemShut
  {NoStop}%
\bibitem [{\citenamefont {Yang}\ \emph {et~al.}(2023)\citenamefont {Yang},
  \citenamefont {Guo}, \citenamefont {Zhang}, \citenamefont {Du}, \citenamefont
  {Zhang}, \citenamefont {Tao}, \citenamefont {Chen}, \citenamefont {Duan},
  \citenamefont {Jia}, \citenamefont {Kong},\ and\ \citenamefont
  {Guo}}]{RN331}%
  \BibitemOpen
  \bibfield  {author} {\bibinfo {author} {\bibfnamefont {X.-X.}\ \bibnamefont
  {Yang}}, \bibinfo {author} {\bibfnamefont {L.-L.}\ \bibnamefont {Guo}},
  \bibinfo {author} {\bibfnamefont {H.-F.}\ \bibnamefont {Zhang}}, \bibinfo
  {author} {\bibfnamefont {L.}~\bibnamefont {Du}}, \bibinfo {author}
  {\bibfnamefont {C.}~\bibnamefont {Zhang}}, \bibinfo {author} {\bibfnamefont
  {H.-R.}\ \bibnamefont {Tao}}, \bibinfo {author} {\bibfnamefont
  {Y.}~\bibnamefont {Chen}}, \bibinfo {author} {\bibfnamefont {P.}~\bibnamefont
  {Duan}}, \bibinfo {author} {\bibfnamefont {Z.-L.}\ \bibnamefont {Jia}},
  \bibinfo {author} {\bibfnamefont {W.-C.}\ \bibnamefont {Kong}},\ and\
  \bibinfo {author} {\bibfnamefont {G.-P.}\ \bibnamefont {Guo}},\ }\href
  {https://doi.org/10.1103/PhysRevApplied.19.044076} {\bibfield  {journal}
  {\bibinfo  {journal} {Physical Review Applied}\ }\textbf {\bibinfo {volume}
  {19}},\ \bibinfo {pages} {044076} (\bibinfo {year} {2023})}\BibitemShut
  {NoStop}%
\bibitem [{\citenamefont {Liu}\ \emph {et~al.}(2023)\citenamefont {Liu},
  \citenamefont {Yan}, \citenamefont {Zhang}, \citenamefont {Yung},
  \citenamefont {Su},\ and\ \citenamefont {Shan}}]{RN361}%
  \BibitemOpen
  \bibfield  {author} {\bibinfo {author} {\bibfnamefont {B.-J.}\ \bibnamefont
  {Liu}}, \bibinfo {author} {\bibfnamefont {L.~L.}\ \bibnamefont {Yan}},
  \bibinfo {author} {\bibfnamefont {Y.}~\bibnamefont {Zhang}}, \bibinfo
  {author} {\bibfnamefont {M.~H.}\ \bibnamefont {Yung}}, \bibinfo {author}
  {\bibfnamefont {S.-L.}\ \bibnamefont {Su}},\ and\ \bibinfo {author}
  {\bibfnamefont {C.~X.}\ \bibnamefont {Shan}},\ }\href
  {https://doi.org/10.1103/PhysRevResearch.5.013059} {\bibfield  {journal}
  {\bibinfo  {journal} {Physical Review Research}\ }\textbf {\bibinfo {volume}
  {5}},\ \bibinfo {pages} {013059} (\bibinfo {year} {2023})}\BibitemShut
  {NoStop}%
\bibitem [{\citenamefont {Samuel}\ and\ \citenamefont
  {Bhandari}(1988)}]{RN154}%
  \BibitemOpen
  \bibfield  {author} {\bibinfo {author} {\bibfnamefont {J.}~\bibnamefont
  {Samuel}}\ and\ \bibinfo {author} {\bibfnamefont {R.}~\bibnamefont
  {Bhandari}},\ }\href {https://doi.org/10.1103/PhysRevLett.60.2339} {\bibfield
   {journal} {\bibinfo  {journal} {Physical Review Letters}\ }\textbf {\bibinfo
  {volume} {60}},\ \bibinfo {pages} {2339} (\bibinfo {year}
  {1988})}\BibitemShut {NoStop}%
\bibitem [{\citenamefont {Anandan}\ and\ \citenamefont
  {Aharonov}(1990)}]{RN136}%
  \BibitemOpen
  \bibfield  {author} {\bibinfo {author} {\bibfnamefont {J.}~\bibnamefont
  {Anandan}}\ and\ \bibinfo {author} {\bibfnamefont {Y.}~\bibnamefont
  {Aharonov}},\ }\href {https://doi.org/10.1103/PhysRevLett.65.1697} {\bibfield
   {journal} {\bibinfo  {journal} {Physical review letters}\ }\textbf {\bibinfo
  {volume} {65}},\ \bibinfo {pages} {1697} (\bibinfo {year}
  {1990})}\BibitemShut {NoStop}%
\bibitem [{\citenamefont {Anandan}(1991)}]{RN138}%
  \BibitemOpen
  \bibfield  {author} {\bibinfo {author} {\bibfnamefont {J.}~\bibnamefont
  {Anandan}},\ }\href {https://doi.org/10.1007/BF00732829} {\bibfield
  {journal} {\bibinfo  {journal} {Foundations of Physics}\ }\textbf {\bibinfo
  {volume} {21}},\ \bibinfo {pages} {1265} (\bibinfo {year}
  {1991})}\BibitemShut {NoStop}%
\bibitem [{\citenamefont {Levitin}\ and\ \citenamefont
  {Toffoli}(2009)}]{RN286}%
  \BibitemOpen
  \bibfield  {author} {\bibinfo {author} {\bibfnamefont {L.~B.}\ \bibnamefont
  {Levitin}}\ and\ \bibinfo {author} {\bibfnamefont {T.}~\bibnamefont
  {Toffoli}},\ }\href {https://doi.org/10.1103/PhysRevLett.103.160502}
  {\bibfield  {journal} {\bibinfo  {journal} {Physical Review Letters}\
  }\textbf {\bibinfo {volume} {103}},\ \bibinfo {pages} {160502} (\bibinfo
  {year} {2009})}\BibitemShut {NoStop}%
\bibitem [{\citenamefont {Campaioli}\ \emph {et~al.}(2018)\citenamefont
  {Campaioli}, \citenamefont {Pollock}, \citenamefont {Binder},\ and\
  \citenamefont {Modi}}]{RN291}%
  \BibitemOpen
  \bibfield  {author} {\bibinfo {author} {\bibfnamefont {F.}~\bibnamefont
  {Campaioli}}, \bibinfo {author} {\bibfnamefont {F.~A.}\ \bibnamefont
  {Pollock}}, \bibinfo {author} {\bibfnamefont {F.~C.}\ \bibnamefont
  {Binder}},\ and\ \bibinfo {author} {\bibfnamefont {K.}~\bibnamefont {Modi}},\
  }\href {https://doi.org/10.1103/PhysRevLett.120.060409} {\bibfield  {journal}
  {\bibinfo  {journal} {Physical Review Letters}\ }\textbf {\bibinfo {volume}
  {120}},\ \bibinfo {pages} {060409} (\bibinfo {year} {2018})}\BibitemShut
  {NoStop}%
\bibitem [{\citenamefont {del Campo}(2021)}]{RN290}%
  \BibitemOpen
  \bibfield  {author} {\bibinfo {author} {\bibfnamefont {A.}~\bibnamefont {del
  Campo}},\ }\href {https://doi.org/10.1103/PhysRevLett.126.180603} {\bibfield
  {journal} {\bibinfo  {journal} {Physical Review Letters}\ }\textbf {\bibinfo
  {volume} {126}},\ \bibinfo {pages} {180603} (\bibinfo {year}
  {2021})}\BibitemShut {NoStop}%
\bibitem [{\citenamefont {Sun}\ \emph {et~al.}(2021)\citenamefont {Sun},
  \citenamefont {Peng}, \citenamefont {Hu},\ and\ \citenamefont
  {Zheng}}]{RN292}%
  \BibitemOpen
  \bibfield  {author} {\bibinfo {author} {\bibfnamefont {S.}~\bibnamefont
  {Sun}}, \bibinfo {author} {\bibfnamefont {Y.}~\bibnamefont {Peng}}, \bibinfo
  {author} {\bibfnamefont {X.}~\bibnamefont {Hu}},\ and\ \bibinfo {author}
  {\bibfnamefont {Y.}~\bibnamefont {Zheng}},\ }\href
  {https://doi.org/10.1103/PhysRevLett.127.100404} {\bibfield  {journal}
  {\bibinfo  {journal} {Physical Review Letters}\ }\textbf {\bibinfo {volume}
  {127}},\ \bibinfo {pages} {100404} (\bibinfo {year} {2021})}\BibitemShut
  {NoStop}%
\bibitem [{\citenamefont {Margolus}\ and\ \citenamefont
  {Levitin}(1998)}]{RN295}%
  \BibitemOpen
  \bibfield  {author} {\bibinfo {author} {\bibfnamefont {N.}~\bibnamefont
  {Margolus}}\ and\ \bibinfo {author} {\bibfnamefont {L.~B.}\ \bibnamefont
  {Levitin}},\ }\href
  {https://doi.org/https://doi.org/10.1016/S0167-2789(98)00054-2} {\bibfield
  {journal} {\bibinfo  {journal} {Physica D: Nonlinear Phenomena}\ }\textbf
  {\bibinfo {volume} {120}},\ \bibinfo {pages} {188} (\bibinfo {year}
  {1998})}\BibitemShut {NoStop}%
\bibitem [{\citenamefont {Lloyd}(2000)}]{RN294}%
  \BibitemOpen
  \bibfield  {author} {\bibinfo {author} {\bibfnamefont {S.}~\bibnamefont
  {Lloyd}},\ }\href {https://doi.org/10.1038/35023282} {\bibfield  {journal}
  {\bibinfo  {journal} {Nature}\ }\textbf {\bibinfo {volume} {406}},\ \bibinfo
  {pages} {1047} (\bibinfo {year} {2000})}\BibitemShut {NoStop}%
\bibitem [{\citenamefont {Santos}\ and\ \citenamefont {Sarandy}(2015)}]{RN297}%
  \BibitemOpen
  \bibfield  {author} {\bibinfo {author} {\bibfnamefont {A.~C.}\ \bibnamefont
  {Santos}}\ and\ \bibinfo {author} {\bibfnamefont {M.~S.}\ \bibnamefont
  {Sarandy}},\ }\href {https://doi.org/10.1038/srep15775} {\bibfield  {journal}
  {\bibinfo  {journal} {Scientific Reports}\ }\textbf {\bibinfo {volume} {5}},\
  \bibinfo {pages} {15775} (\bibinfo {year} {2015})}\BibitemShut {NoStop}%
\bibitem [{\citenamefont {Jordan}(2017)}]{RN296}%
  \BibitemOpen
  \bibfield  {author} {\bibinfo {author} {\bibfnamefont {S.~P.}\ \bibnamefont
  {Jordan}},\ }\href {https://doi.org/10.1103/PhysRevA.95.032305} {\bibfield
  {journal} {\bibinfo  {journal} {Physical Review A}\ }\textbf {\bibinfo
  {volume} {95}},\ \bibinfo {pages} {032305} (\bibinfo {year}
  {2017})}\BibitemShut {NoStop}%
\bibitem [{\citenamefont {Braunstein}(1992)}]{RN293}%
  \BibitemOpen
  \bibfield  {author} {\bibinfo {author} {\bibfnamefont {S.~L.}\ \bibnamefont
  {Braunstein}},\ }\href {https://doi.org/10.1103/PhysRevLett.69.3598}
  {\bibfield  {journal} {\bibinfo  {journal} {Physical Review Letters}\
  }\textbf {\bibinfo {volume} {69}},\ \bibinfo {pages} {3598} (\bibinfo {year}
  {1992})}\BibitemShut {NoStop}%
\bibitem [{\citenamefont {Giovannetti}\ \emph {et~al.}(2006)\citenamefont
  {Giovannetti}, \citenamefont {Lloyd},\ and\ \citenamefont {Maccone}}]{RN298}%
  \BibitemOpen
  \bibfield  {author} {\bibinfo {author} {\bibfnamefont {V.}~\bibnamefont
  {Giovannetti}}, \bibinfo {author} {\bibfnamefont {S.}~\bibnamefont {Lloyd}},\
  and\ \bibinfo {author} {\bibfnamefont {L.}~\bibnamefont {Maccone}},\ }\href
  {https://doi.org/10.1103/PhysRevLett.96.010401} {\bibfield  {journal}
  {\bibinfo  {journal} {Physical Review Letters}\ }\textbf {\bibinfo {volume}
  {96}},\ \bibinfo {pages} {010401} (\bibinfo {year} {2006})}\BibitemShut
  {NoStop}%
\bibitem [{\citenamefont {Zwierz}\ \emph {et~al.}(2010)\citenamefont {Zwierz},
  \citenamefont {Pérez-Delgado},\ and\ \citenamefont {Kok}}]{RN300}%
  \BibitemOpen
  \bibfield  {author} {\bibinfo {author} {\bibfnamefont {M.}~\bibnamefont
  {Zwierz}}, \bibinfo {author} {\bibfnamefont {C.~A.}\ \bibnamefont
  {Pérez-Delgado}},\ and\ \bibinfo {author} {\bibfnamefont {P.}~\bibnamefont
  {Kok}},\ }\href {https://doi.org/10.1103/PhysRevLett.105.180402} {\bibfield
  {journal} {\bibinfo  {journal} {Physical Review Letters}\ }\textbf {\bibinfo
  {volume} {105}},\ \bibinfo {pages} {180402} (\bibinfo {year}
  {2010})}\BibitemShut {NoStop}%
\bibitem [{\citenamefont {Giovannetti}\ \emph {et~al.}(2011)\citenamefont
  {Giovannetti}, \citenamefont {Lloyd},\ and\ \citenamefont {Maccone}}]{RN299}%
  \BibitemOpen
  \bibfield  {author} {\bibinfo {author} {\bibfnamefont {V.}~\bibnamefont
  {Giovannetti}}, \bibinfo {author} {\bibfnamefont {S.}~\bibnamefont {Lloyd}},\
  and\ \bibinfo {author} {\bibfnamefont {L.}~\bibnamefont {Maccone}},\ }\href
  {https://doi.org/10.1038/nphoton.2011.35} {\bibfield  {journal} {\bibinfo
  {journal} {Nature Photonics}\ }\textbf {\bibinfo {volume} {5}},\ \bibinfo
  {pages} {222} (\bibinfo {year} {2011})}\BibitemShut {NoStop}%
\bibitem [{\citenamefont {Walmsley}\ and\ \citenamefont
  {Rabitz}(2003)}]{RN301}%
  \BibitemOpen
  \bibfield  {author} {\bibinfo {author} {\bibfnamefont {I.}~\bibnamefont
  {Walmsley}}\ and\ \bibinfo {author} {\bibfnamefont {E.}~\bibnamefont
  {Rabitz}},\ }\href {https://doi.org/Doi 10.1063/1.1611352} {\bibfield
  {journal} {\bibinfo  {journal} {Physics Today}\ }\textbf {\bibinfo {volume}
  {56}},\ \bibinfo {pages} {43} (\bibinfo {year} {2003})}\BibitemShut {NoStop}%
\bibitem [{\citenamefont {Carlini}\ \emph {et~al.}(2006)\citenamefont
  {Carlini}, \citenamefont {Hosoya}, \citenamefont {Koike},\ and\ \citenamefont
  {Okudaira}}]{RN287}%
  \BibitemOpen
  \bibfield  {author} {\bibinfo {author} {\bibfnamefont {A.}~\bibnamefont
  {Carlini}}, \bibinfo {author} {\bibfnamefont {A.}~\bibnamefont {Hosoya}},
  \bibinfo {author} {\bibfnamefont {T.}~\bibnamefont {Koike}},\ and\ \bibinfo
  {author} {\bibfnamefont {Y.}~\bibnamefont {Okudaira}},\ }\href
  {https://doi.org/10.1103/PhysRevLett.96.060503} {\bibfield  {journal}
  {\bibinfo  {journal} {Physical Review Letters}\ }\textbf {\bibinfo {volume}
  {96}},\ \bibinfo {pages} {060503} (\bibinfo {year} {2006})}\BibitemShut
  {NoStop}%
\bibitem [{\citenamefont {Deffner}\ and\ \citenamefont
  {Campbell}(2017)}]{RN281}%
  \BibitemOpen
  \bibfield  {author} {\bibinfo {author} {\bibfnamefont {S.}~\bibnamefont
  {Deffner}}\ and\ \bibinfo {author} {\bibfnamefont {S.}~\bibnamefont
  {Campbell}},\ }\href {https://doi.org/10.1088/1751-8121/aa86c6} {\bibfield
  {journal} {\bibinfo  {journal} {Journal of Physics A: Mathematical and
  Theoretical}\ }\textbf {\bibinfo {volume} {50}},\ \bibinfo {pages} {453001}
  (\bibinfo {year} {2017})}\BibitemShut {NoStop}%
\bibitem [{\citenamefont {Garcia}\ \emph {et~al.}(2022)\citenamefont {Garcia},
  \citenamefont {Bofill}, \citenamefont {Moreira},\ and\ \citenamefont
  {Albareda}}]{RN285}%
  \BibitemOpen
  \bibfield  {author} {\bibinfo {author} {\bibfnamefont {L.}~\bibnamefont
  {Garcia}}, \bibinfo {author} {\bibfnamefont {J.~M.}\ \bibnamefont {Bofill}},
  \bibinfo {author} {\bibfnamefont {I.~D.~R.}\ \bibnamefont {Moreira}},\ and\
  \bibinfo {author} {\bibfnamefont {G.}~\bibnamefont {Albareda}},\ }\href
  {https://doi.org/ARTN 180402 10.1103/PhysRevLett.129.180402} {\bibfield
  {journal} {\bibinfo  {journal} {Physical Review Letters}\ }\textbf {\bibinfo
  {volume} {129}},\ \bibinfo {pages} {180402} (\bibinfo {year}
  {2022})}\BibitemShut {NoStop}%
\bibitem [{\citenamefont {Rabi}(1936)}]{RN261}%
  \BibitemOpen
  \bibfield  {author} {\bibinfo {author} {\bibfnamefont {I.~I.}\ \bibnamefont
  {Rabi}},\ }\href {https://doi.org/10.1103/PhysRev.49.324} {\bibfield
  {journal} {\bibinfo  {journal} {Physical Review}\ }\textbf {\bibinfo {volume}
  {49}},\ \bibinfo {pages} {324} (\bibinfo {year} {1936})}\BibitemShut
  {NoStop}%
\bibitem [{\citenamefont {Rabi}(1937)}]{RN262}%
  \BibitemOpen
  \bibfield  {author} {\bibinfo {author} {\bibfnamefont {I.~I.}\ \bibnamefont
  {Rabi}},\ }\href {https://doi.org/10.1103/PhysRev.51.652} {\bibfield
  {journal} {\bibinfo  {journal} {Physical Review}\ }\textbf {\bibinfo {volume}
  {51}},\ \bibinfo {pages} {652} (\bibinfo {year} {1937})}\BibitemShut
  {NoStop}%
\bibitem [{\citenamefont {Shirley}(1965)}]{RN149}%
  \BibitemOpen
  \bibfield  {author} {\bibinfo {author} {\bibfnamefont {J.~H.}\ \bibnamefont
  {Shirley}},\ }\href {https://doi.org/10.1103/PhysRev.138.B979} {\bibfield
  {journal} {\bibinfo  {journal} {Physical Review}\ }\textbf {\bibinfo {volume}
  {138}},\ \bibinfo {pages} {B979} (\bibinfo {year} {1965})}\BibitemShut
  {NoStop}%
\bibitem [{\citenamefont {Milonni}(2019)}]{RN341}%
  \BibitemOpen
  \bibfield  {author} {\bibinfo {author} {\bibfnamefont {P.~W.}\ \bibnamefont
  {Milonni}},\ }\href@noop {} {\emph {\bibinfo {title} {An introduction to
  quantum optics and quantum fluctuations}}}\ (\bibinfo  {publisher} {Oxford
  University Press},\ \bibinfo {year} {2019})\BibitemShut {NoStop}%
\bibitem [{\citenamefont {Leggett}\ \emph {et~al.}(1987)\citenamefont
  {Leggett}, \citenamefont {Chakravarty}, \citenamefont {Dorsey}, \citenamefont
  {Fisher}, \citenamefont {Garg},\ and\ \citenamefont {Zwerger}}]{RN348}%
  \BibitemOpen
  \bibfield  {author} {\bibinfo {author} {\bibfnamefont {A.~J.}\ \bibnamefont
  {Leggett}}, \bibinfo {author} {\bibfnamefont {S.}~\bibnamefont
  {Chakravarty}}, \bibinfo {author} {\bibfnamefont {A.~T.}\ \bibnamefont
  {Dorsey}}, \bibinfo {author} {\bibfnamefont {M.~P.~A.}\ \bibnamefont
  {Fisher}}, \bibinfo {author} {\bibfnamefont {A.}~\bibnamefont {Garg}},\ and\
  \bibinfo {author} {\bibfnamefont {W.}~\bibnamefont {Zwerger}},\ }\href
  {https://doi.org/10.1103/RevModPhys.59.1} {\bibfield  {journal} {\bibinfo
  {journal} {Reviews of Modern Physics}\ }\textbf {\bibinfo {volume} {59}},\
  \bibinfo {pages} {1} (\bibinfo {year} {1987})}\BibitemShut {NoStop}%
\bibitem [{\citenamefont {Dakhnovskii}\ and\ \citenamefont
  {Bavli}(1993)}]{RN342}%
  \BibitemOpen
  \bibfield  {author} {\bibinfo {author} {\bibfnamefont {Y.}~\bibnamefont
  {Dakhnovskii}}\ and\ \bibinfo {author} {\bibfnamefont {R.}~\bibnamefont
  {Bavli}},\ }\href {https://doi.org/10.1103/PhysRevB.48.11020} {\bibfield
  {journal} {\bibinfo  {journal} {Physical Review B}\ }\textbf {\bibinfo
  {volume} {48}},\ \bibinfo {pages} {11020} (\bibinfo {year}
  {1993})}\BibitemShut {NoStop}%
\bibitem [{\citenamefont {Dakhnovskii}\ and\ \citenamefont
  {Metiu}(1993)}]{RN344}%
  \BibitemOpen
  \bibfield  {author} {\bibinfo {author} {\bibfnamefont {Y.}~\bibnamefont
  {Dakhnovskii}}\ and\ \bibinfo {author} {\bibfnamefont {H.}~\bibnamefont
  {Metiu}},\ }\href {https://doi.org/10.1103/PhysRevA.48.2342} {\bibfield
  {journal} {\bibinfo  {journal} {Physical Review A}\ }\textbf {\bibinfo
  {volume} {48}},\ \bibinfo {pages} {2342} (\bibinfo {year}
  {1993})}\BibitemShut {NoStop}%
\bibitem [{\citenamefont {Rudner}\ \emph {et~al.}(2008)\citenamefont {Rudner},
  \citenamefont {Shytov}, \citenamefont {Levitov}, \citenamefont {Berns},
  \citenamefont {Oliver}, \citenamefont {Valenzuela},\ and\ \citenamefont
  {Orlando}}]{RN343}%
  \BibitemOpen
  \bibfield  {author} {\bibinfo {author} {\bibfnamefont {M.~S.}\ \bibnamefont
  {Rudner}}, \bibinfo {author} {\bibfnamefont {A.~V.}\ \bibnamefont {Shytov}},
  \bibinfo {author} {\bibfnamefont {L.~S.}\ \bibnamefont {Levitov}}, \bibinfo
  {author} {\bibfnamefont {D.~M.}\ \bibnamefont {Berns}}, \bibinfo {author}
  {\bibfnamefont {W.~D.}\ \bibnamefont {Oliver}}, \bibinfo {author}
  {\bibfnamefont {S.~O.}\ \bibnamefont {Valenzuela}},\ and\ \bibinfo {author}
  {\bibfnamefont {T.~P.}\ \bibnamefont {Orlando}},\ }\href
  {https://doi.org/10.1103/PhysRevLett.101.190502} {\bibfield  {journal}
  {\bibinfo  {journal} {Physical Review Letters}\ }\textbf {\bibinfo {volume}
  {101}},\ \bibinfo {pages} {190502} (\bibinfo {year} {2008})}\BibitemShut
  {NoStop}%
\bibitem [{\citenamefont {Ramsay}\ \emph {et~al.}(2010)\citenamefont {Ramsay},
  \citenamefont {Godden}, \citenamefont {Boyle}, \citenamefont {Gauger},
  \citenamefont {Nazir}, \citenamefont {Lovett}, \citenamefont {Fox},\ and\
  \citenamefont {Skolnick}}]{RN346}%
  \BibitemOpen
  \bibfield  {author} {\bibinfo {author} {\bibfnamefont {A.~J.}\ \bibnamefont
  {Ramsay}}, \bibinfo {author} {\bibfnamefont {T.~M.}\ \bibnamefont {Godden}},
  \bibinfo {author} {\bibfnamefont {S.~J.}\ \bibnamefont {Boyle}}, \bibinfo
  {author} {\bibfnamefont {E.~M.}\ \bibnamefont {Gauger}}, \bibinfo {author}
  {\bibfnamefont {A.}~\bibnamefont {Nazir}}, \bibinfo {author} {\bibfnamefont
  {B.~W.}\ \bibnamefont {Lovett}}, \bibinfo {author} {\bibfnamefont {A.~M.}\
  \bibnamefont {Fox}},\ and\ \bibinfo {author} {\bibfnamefont {M.~S.}\
  \bibnamefont {Skolnick}},\ }\href
  {https://doi.org/10.1103/PhysRevLett.105.177402} {\bibfield  {journal}
  {\bibinfo  {journal} {Physical Review Letters}\ }\textbf {\bibinfo {volume}
  {105}},\ \bibinfo {pages} {177402} (\bibinfo {year} {2010})}\BibitemShut
  {NoStop}%
\bibitem [{\citenamefont {Destefani}\ \emph {et~al.}(2022)\citenamefont
  {Destefani}, \citenamefont {Villani}, \citenamefont {Cartoixà},
  \citenamefont {Feiginov},\ and\ \citenamefont {Oriols}}]{RN330}%
  \BibitemOpen
  \bibfield  {author} {\bibinfo {author} {\bibfnamefont {C.~F.}\ \bibnamefont
  {Destefani}}, \bibinfo {author} {\bibfnamefont {M.}~\bibnamefont {Villani}},
  \bibinfo {author} {\bibfnamefont {X.}~\bibnamefont {Cartoixà}}, \bibinfo
  {author} {\bibfnamefont {M.}~\bibnamefont {Feiginov}},\ and\ \bibinfo
  {author} {\bibfnamefont {X.}~\bibnamefont {Oriols}},\ }\href
  {https://doi.org/10.1103/PhysRevB.106.205306} {\bibfield  {journal} {\bibinfo
   {journal} {Physical Review B}\ }\textbf {\bibinfo {volume} {106}},\ \bibinfo
  {pages} {205306} (\bibinfo {year} {2022})}\BibitemShut {NoStop}%
\bibitem [{\citenamefont {Manson}\ \emph {et~al.}(2016)\citenamefont {Manson},
  \citenamefont {Roy-Choudhury},\ and\ \citenamefont {Hughes}}]{RN347}%
  \BibitemOpen
  \bibfield  {author} {\bibinfo {author} {\bibfnamefont {R.}~\bibnamefont
  {Manson}}, \bibinfo {author} {\bibfnamefont {K.}~\bibnamefont
  {Roy-Choudhury}},\ and\ \bibinfo {author} {\bibfnamefont {S.}~\bibnamefont
  {Hughes}},\ }\href {https://doi.org/10.1103/PhysRevB.93.155423} {\bibfield
  {journal} {\bibinfo  {journal} {Physical Review B}\ }\textbf {\bibinfo
  {volume} {93}},\ \bibinfo {pages} {155423} (\bibinfo {year}
  {2016})}\BibitemShut {NoStop}%
\bibitem [{\citenamefont {Vion}\ \emph {et~al.}(2002)\citenamefont {Vion},
  \citenamefont {Aassime}, \citenamefont {Cottet}, \citenamefont {Joyez},
  \citenamefont {Pothier}, \citenamefont {Urbina}, \citenamefont {Esteve},\
  and\ \citenamefont {Devoret}}]{RN309}%
  \BibitemOpen
  \bibfield  {author} {\bibinfo {author} {\bibfnamefont {D.}~\bibnamefont
  {Vion}}, \bibinfo {author} {\bibfnamefont {A.}~\bibnamefont {Aassime}},
  \bibinfo {author} {\bibfnamefont {A.}~\bibnamefont {Cottet}}, \bibinfo
  {author} {\bibfnamefont {P.}~\bibnamefont {Joyez}}, \bibinfo {author}
  {\bibfnamefont {H.}~\bibnamefont {Pothier}}, \bibinfo {author} {\bibfnamefont
  {C.}~\bibnamefont {Urbina}}, \bibinfo {author} {\bibfnamefont
  {D.}~\bibnamefont {Esteve}},\ and\ \bibinfo {author} {\bibfnamefont {M.~H.}\
  \bibnamefont {Devoret}},\ }\href
  {https://doi.org/doi:10.1126/science.1069372} {\bibfield  {journal} {\bibinfo
   {journal} {Science}\ }\textbf {\bibinfo {volume} {296}},\ \bibinfo {pages}
  {886} (\bibinfo {year} {2002})}\BibitemShut {NoStop}%
\bibitem [{\citenamefont {Il’ichev}\ \emph {et~al.}(2003)\citenamefont
  {Il’ichev}, \citenamefont {Oukhanski}, \citenamefont {Izmalkov},
  \citenamefont {Wagner}, \citenamefont {Grajcar}, \citenamefont {Meyer},
  \citenamefont {Smirnov}, \citenamefont {Maassen van~den Brink}, \citenamefont
  {Amin},\ and\ \citenamefont {Zagoskin}}]{RN340}%
  \BibitemOpen
  \bibfield  {author} {\bibinfo {author} {\bibfnamefont {E.}~\bibnamefont
  {Il’ichev}}, \bibinfo {author} {\bibfnamefont {N.}~\bibnamefont
  {Oukhanski}}, \bibinfo {author} {\bibfnamefont {A.}~\bibnamefont {Izmalkov}},
  \bibinfo {author} {\bibfnamefont {T.}~\bibnamefont {Wagner}}, \bibinfo
  {author} {\bibfnamefont {M.}~\bibnamefont {Grajcar}}, \bibinfo {author}
  {\bibfnamefont {H.~G.}\ \bibnamefont {Meyer}}, \bibinfo {author}
  {\bibfnamefont {A.~Y.}\ \bibnamefont {Smirnov}}, \bibinfo {author}
  {\bibfnamefont {A.}~\bibnamefont {Maassen van~den Brink}}, \bibinfo {author}
  {\bibfnamefont {M.~H.~S.}\ \bibnamefont {Amin}},\ and\ \bibinfo {author}
  {\bibfnamefont {A.~M.}\ \bibnamefont {Zagoskin}},\ }\href
  {https://doi.org/10.1103/PhysRevLett.91.097906} {\bibfield  {journal}
  {\bibinfo  {journal} {Physical Review Letters}\ }\textbf {\bibinfo {volume}
  {91}},\ \bibinfo {pages} {097906} (\bibinfo {year} {2003})}\BibitemShut
  {NoStop}%
\bibitem [{\citenamefont {Deng}\ \emph {et~al.}(2015)\citenamefont {Deng},
  \citenamefont {Orgiazzi}, \citenamefont {Shen}, \citenamefont {Ashhab},\ and\
  \citenamefont {Lupascu}}]{RN339}%
  \BibitemOpen
  \bibfield  {author} {\bibinfo {author} {\bibfnamefont {C.}~\bibnamefont
  {Deng}}, \bibinfo {author} {\bibfnamefont {J.-L.}\ \bibnamefont {Orgiazzi}},
  \bibinfo {author} {\bibfnamefont {F.}~\bibnamefont {Shen}}, \bibinfo {author}
  {\bibfnamefont {S.}~\bibnamefont {Ashhab}},\ and\ \bibinfo {author}
  {\bibfnamefont {A.}~\bibnamefont {Lupascu}},\ }\href
  {https://doi.org/10.1103/PhysRevLett.115.133601} {\bibfield  {journal}
  {\bibinfo  {journal} {Physical Review Letters}\ }\textbf {\bibinfo {volume}
  {115}},\ \bibinfo {pages} {133601} (\bibinfo {year} {2015})}\BibitemShut
  {NoStop}%
\bibitem [{\citenamefont {Grossmann}\ \emph {et~al.}(1991)\citenamefont
  {Grossmann}, \citenamefont {Dittrich}, \citenamefont {Jung},\ and\
  \citenamefont {Hänggi}}]{RN268}%
  \BibitemOpen
  \bibfield  {author} {\bibinfo {author} {\bibfnamefont {F.}~\bibnamefont
  {Grossmann}}, \bibinfo {author} {\bibfnamefont {T.}~\bibnamefont {Dittrich}},
  \bibinfo {author} {\bibfnamefont {P.}~\bibnamefont {Jung}},\ and\ \bibinfo
  {author} {\bibfnamefont {P.}~\bibnamefont {Hänggi}},\ }\href
  {https://doi.org/10.1103/PhysRevLett.67.516} {\bibfield  {journal} {\bibinfo
  {journal} {Physical Review Letters}\ }\textbf {\bibinfo {volume} {67}},\
  \bibinfo {pages} {516} (\bibinfo {year} {1991})}\BibitemShut {NoStop}%
\bibitem [{\citenamefont {Goychuk}\ and\ \citenamefont
  {Hänggi}(2005)}]{RN269}%
  \BibitemOpen
  \bibfield  {author} {\bibinfo {author} {\bibfnamefont {I.}~\bibnamefont
  {Goychuk}}\ and\ \bibinfo {author} {\bibfnamefont {P.}~\bibnamefont
  {Hänggi}},\ }\href {https://doi.org/10.1080/00018730500429701} {\bibfield
  {journal} {\bibinfo  {journal} {Advances in Physics}\ }\textbf {\bibinfo
  {volume} {54}},\ \bibinfo {pages} {525} (\bibinfo {year} {2005})}\BibitemShut
  {NoStop}%
\bibitem [{\citenamefont {Chen}\ \emph {et~al.}(2020)\citenamefont {Chen},
  \citenamefont {L\"u}, \citenamefont {Yan},\ and\ \citenamefont
  {Zheng}}]{RN345}%
  \BibitemOpen
  \bibfield  {author} {\bibinfo {author} {\bibfnamefont {Y.}~\bibnamefont
  {Chen}}, \bibinfo {author} {\bibfnamefont {Z.}~\bibnamefont {L\"u}}, \bibinfo
  {author} {\bibfnamefont {Y.}~\bibnamefont {Yan}},\ and\ \bibinfo {author}
  {\bibfnamefont {H.}~\bibnamefont {Zheng}},\ }\href
  {https://doi.org/10.1103/PhysRevA.102.053703} {\bibfield  {journal} {\bibinfo
   {journal} {Physical Review A}\ }\textbf {\bibinfo {volume} {102}},\ \bibinfo
  {pages} {053703} (\bibinfo {year} {2020})}\BibitemShut {NoStop}%
\bibitem [{\citenamefont {Braak}(2011)}]{RN185}%
  \BibitemOpen
  \bibfield  {author} {\bibinfo {author} {\bibfnamefont {D.}~\bibnamefont
  {Braak}},\ }\href {https://doi.org/10.1103/PhysRevLett.107.100401} {\bibfield
   {journal} {\bibinfo  {journal} {Physical Review Letters}\ }\textbf {\bibinfo
  {volume} {107}},\ \bibinfo {pages} {100401} (\bibinfo {year}
  {2011})}\BibitemShut {NoStop}%
\bibitem [{\citenamefont {Yoshihara}\ \emph {et~al.}(2014)\citenamefont
  {Yoshihara}, \citenamefont {Nakamura}, \citenamefont {Yan}, \citenamefont
  {Gustavsson}, \citenamefont {Bylander}, \citenamefont {Oliver},\ and\
  \citenamefont {Tsai}}]{RN258}%
  \BibitemOpen
  \bibfield  {author} {\bibinfo {author} {\bibfnamefont {F.}~\bibnamefont
  {Yoshihara}}, \bibinfo {author} {\bibfnamefont {Y.}~\bibnamefont {Nakamura}},
  \bibinfo {author} {\bibfnamefont {F.}~\bibnamefont {Yan}}, \bibinfo {author}
  {\bibfnamefont {S.}~\bibnamefont {Gustavsson}}, \bibinfo {author}
  {\bibfnamefont {J.}~\bibnamefont {Bylander}}, \bibinfo {author}
  {\bibfnamefont {W.~D.}\ \bibnamefont {Oliver}},\ and\ \bibinfo {author}
  {\bibfnamefont {J.-S.}\ \bibnamefont {Tsai}},\ }\href
  {https://doi.org/10.1103/PhysRevB.89.020503} {\bibfield  {journal} {\bibinfo
  {journal} {Physical Review B}\ }\textbf {\bibinfo {volume} {89}},\ \bibinfo
  {pages} {020503} (\bibinfo {year} {2014})}\BibitemShut {NoStop}%
\bibitem [{\citenamefont {Hausinger}\ and\ \citenamefont
  {Grifoni}(2010)}]{RN147}%
  \BibitemOpen
  \bibfield  {author} {\bibinfo {author} {\bibfnamefont {J.}~\bibnamefont
  {Hausinger}}\ and\ \bibinfo {author} {\bibfnamefont {M.}~\bibnamefont
  {Grifoni}},\ }\href {https://doi.org/10.1103/PhysRevA.81.022117} {\bibfield
  {journal} {\bibinfo  {journal} {Physical Review A}\ }\textbf {\bibinfo
  {volume} {81}},\ \bibinfo {pages} {022117} (\bibinfo {year}
  {2010})}\BibitemShut {NoStop}%
\bibitem [{\citenamefont {Lü}\ \emph {et~al.}(2016)\citenamefont {Lü},
  \citenamefont {Yan}, \citenamefont {Goan},\ and\ \citenamefont
  {Zheng}}]{RN103}%
  \BibitemOpen
  \bibfield  {author} {\bibinfo {author} {\bibfnamefont {Z.}~\bibnamefont
  {Lü}}, \bibinfo {author} {\bibfnamefont {Y.}~\bibnamefont {Yan}}, \bibinfo
  {author} {\bibfnamefont {H.-S.}\ \bibnamefont {Goan}},\ and\ \bibinfo
  {author} {\bibfnamefont {H.}~\bibnamefont {Zheng}},\ }\href@noop {}
  {\bibfield  {journal} {\bibinfo  {journal} {Physical Review A}\ }\textbf
  {\bibinfo {volume} {93}},\ \bibinfo {pages} {033803} (\bibinfo {year}
  {2016})}\BibitemShut {NoStop}%
\bibitem [{\citenamefont {Liu}\ \emph {et~al.}(2021)\citenamefont {Liu},
  \citenamefont {Lü},\ and\ \citenamefont {Zheng}}]{RN102}%
  \BibitemOpen
  \bibfield  {author} {\bibinfo {author} {\bibfnamefont {S.}~\bibnamefont
  {Liu}}, \bibinfo {author} {\bibfnamefont {Z.}~\bibnamefont {Lü}},\ and\
  \bibinfo {author} {\bibfnamefont {H.}~\bibnamefont {Zheng}},\ }\href
  {https://doi.org/ARTN 445302 10.1088/1751-8121/ac2a04} {\bibfield  {journal}
  {\bibinfo  {journal} {Journal of Physics A: Mathematical and Theoretical}\
  }\textbf {\bibinfo {volume} {54}},\ \bibinfo {pages} {445302} (\bibinfo
  {year} {2021})}\BibitemShut {NoStop}%
\bibitem [{\citenamefont {Seleznyova}(1993)}]{RN168}%
  \BibitemOpen
  \bibfield  {author} {\bibinfo {author} {\bibfnamefont {A.~N.}\ \bibnamefont
  {Seleznyova}},\ }\href {https://doi.org/Doi 10.1088/0305-4470/26/4/025}
  {\bibfield  {journal} {\bibinfo  {journal} {Journal of Physics A}\ }\textbf
  {\bibinfo {volume} {26}},\ \bibinfo {pages} {981} (\bibinfo {year}
  {1993})}\BibitemShut {NoStop}%
\bibitem [{\citenamefont {Oliver}\ \emph {et~al.}(2005)\citenamefont {Oliver},
  \citenamefont {Yu}, \citenamefont {Lee}, \citenamefont {Berggren},
  \citenamefont {Levitov},\ and\ \citenamefont {Orlando}}]{add4}%
  \BibitemOpen
  \bibfield  {author} {\bibinfo {author} {\bibfnamefont {W.~D.}\ \bibnamefont
  {Oliver}}, \bibinfo {author} {\bibfnamefont {Y.}~\bibnamefont {Yu}}, \bibinfo
  {author} {\bibfnamefont {J.~C.}\ \bibnamefont {Lee}}, \bibinfo {author}
  {\bibfnamefont {K.~K.}\ \bibnamefont {Berggren}}, \bibinfo {author}
  {\bibfnamefont {L.~S.}\ \bibnamefont {Levitov}},\ and\ \bibinfo {author}
  {\bibfnamefont {T.~P.}\ \bibnamefont {Orlando}},\ }\href
  {https://doi.org/10.1126/science.1119678} {\bibfield  {journal} {\bibinfo
  {journal} {Science}\ }\textbf {\bibinfo {volume} {310}},\ \bibinfo {pages}
  {1653} (\bibinfo {year} {2005})}\BibitemShut {NoStop}%
\bibitem [{\citenamefont {Grifoni}\ and\ \citenamefont {Hänggi}(1998)}]{add3}%
  \BibitemOpen
  \bibfield  {author} {\bibinfo {author} {\bibfnamefont {M.}~\bibnamefont
  {Grifoni}}\ and\ \bibinfo {author} {\bibfnamefont {P.}~\bibnamefont
  {Hänggi}},\ }\href
  {https://doi.org/https://doi.org/10.1016/S0370-1573(98)00022-2} {\bibfield
  {journal} {\bibinfo  {journal} {Physics Reports}\ }\textbf {\bibinfo {volume}
  {304}},\ \bibinfo {pages} {229} (\bibinfo {year} {1998})}\BibitemShut
  {NoStop}%
\bibitem [{\citenamefont {Yabuzaki}\ \emph {et~al.}(1974)\citenamefont
  {Yabuzaki}, \citenamefont {Nakayama}, \citenamefont {Murakami},\ and\
  \citenamefont {Ogawa}}]{add1}%
  \BibitemOpen
  \bibfield  {author} {\bibinfo {author} {\bibfnamefont {T.}~\bibnamefont
  {Yabuzaki}}, \bibinfo {author} {\bibfnamefont {S.}~\bibnamefont {Nakayama}},
  \bibinfo {author} {\bibfnamefont {Y.}~\bibnamefont {Murakami}},\ and\
  \bibinfo {author} {\bibfnamefont {T.}~\bibnamefont {Ogawa}},\ }\href
  {https://doi.org/10.1103/PhysRevA.10.1955} {\bibfield  {journal} {\bibinfo
  {journal} {Physical Review A}\ }\textbf {\bibinfo {volume} {10}},\ \bibinfo
  {pages} {1955} (\bibinfo {year} {1974})}\BibitemShut {NoStop}%
\bibitem [{\citenamefont {Tsukada}\ \emph {et~al.}(1973)\citenamefont
  {Tsukada}, \citenamefont {Koyama},\ and\ \citenamefont {Ogawa}}]{add2}%
  \BibitemOpen
  \bibfield  {author} {\bibinfo {author} {\bibfnamefont {N.}~\bibnamefont
  {Tsukada}}, \bibinfo {author} {\bibfnamefont {T.}~\bibnamefont {Koyama}},\
  and\ \bibinfo {author} {\bibfnamefont {T.}~\bibnamefont {Ogawa}},\ }\href
  {https://doi.org/https://doi.org/10.1016/0375-9601(73)90993-6} {\bibfield
  {journal} {\bibinfo  {journal} {Physics Letters A}\ }\textbf {\bibinfo
  {volume} {44}},\ \bibinfo {pages} {501} (\bibinfo {year} {1973})}\BibitemShut
  {NoStop}%
\bibitem [{\citenamefont {Mangazeev}\ \emph {et~al.}(2021)\citenamefont
  {Mangazeev}, \citenamefont {Batchelor},\ and\ \citenamefont
  {Bazhanov}}]{RN256}%
  \BibitemOpen
  \bibfield  {author} {\bibinfo {author} {\bibfnamefont {V.~V.}\ \bibnamefont
  {Mangazeev}}, \bibinfo {author} {\bibfnamefont {M.~T.}\ \bibnamefont
  {Batchelor}},\ and\ \bibinfo {author} {\bibfnamefont {V.~V.}\ \bibnamefont
  {Bazhanov}},\ }\href {https://doi.org/10.1088/1751-8121/abe426} {\bibfield
  {journal} {\bibinfo  {journal} {Journal of Physics A: Mathematical and
  Theoretical}\ }\textbf {\bibinfo {volume} {54}},\ \bibinfo {pages} {12LT01}
  (\bibinfo {year} {2021})}\BibitemShut {NoStop}%
\bibitem [{\citenamefont {Reyes-Bustos}\ \emph {et~al.}(2021)\citenamefont
  {Reyes-Bustos}, \citenamefont {Braak},\ and\ \citenamefont
  {Wakayama}}]{RN196}%
  \BibitemOpen
  \bibfield  {author} {\bibinfo {author} {\bibfnamefont {C.}~\bibnamefont
  {Reyes-Bustos}}, \bibinfo {author} {\bibfnamefont {D.}~\bibnamefont
  {Braak}},\ and\ \bibinfo {author} {\bibfnamefont {M.}~\bibnamefont
  {Wakayama}},\ }\href {https://doi.org/10.1088/1751-8121/ac0508} {\bibfield
  {journal} {\bibinfo  {journal} {Journal of Physics A: Mathematical and
  Theoretical}\ }\textbf {\bibinfo {volume} {54}},\ \bibinfo {pages} {285202}
  (\bibinfo {year} {2021})}\BibitemShut {NoStop}%
\bibitem [{\citenamefont {Xie}\ and\ \citenamefont {Chen}(2022)}]{RN193}%
  \BibitemOpen
  \bibfield  {author} {\bibinfo {author} {\bibfnamefont {Y.-F.}\ \bibnamefont
  {Xie}}\ and\ \bibinfo {author} {\bibfnamefont {Q.-H.}\ \bibnamefont {Chen}},\
  }\href {https://doi.org/10.1088/1751-8121/ac6842} {\bibfield  {journal}
  {\bibinfo  {journal} {Journal of Physics A: Mathematical and Theoretical}\
  }\textbf {\bibinfo {volume} {55}},\ \bibinfo {pages} {225306} (\bibinfo
  {year} {2022})}\BibitemShut {NoStop}%
\bibitem [{\citenamefont {Irish}\ and\ \citenamefont {Armour}(2022)}]{RN197}%
  \BibitemOpen
  \bibfield  {author} {\bibinfo {author} {\bibfnamefont {E.~K.}\ \bibnamefont
  {Irish}}\ and\ \bibinfo {author} {\bibfnamefont {A.~D.}\ \bibnamefont
  {Armour}},\ }\href {https://doi.org/ARTN 183603
  10.1103/PhysRevLett.129.183603} {\bibfield  {journal} {\bibinfo  {journal}
  {Physical Review Letters}\ }\textbf {\bibinfo {volume} {129}},\ \bibinfo
  {pages} {183603} (\bibinfo {year} {2022})}\BibitemShut {NoStop}%
\bibitem [{\citenamefont {Ashhab}(2017)}]{RN253}%
  \BibitemOpen
  \bibfield  {author} {\bibinfo {author} {\bibfnamefont {S.}~\bibnamefont
  {Ashhab}},\ }\href {https://doi.org/10.1088/1751-8121/aa5f6e} {\bibfield
  {journal} {\bibinfo  {journal} {Journal of Physics A: Mathematical and
  Theoretical}\ }\textbf {\bibinfo {volume} {50}},\ \bibinfo {pages} {134002}
  (\bibinfo {year} {2017})}\BibitemShut {NoStop}%
\end{thebibliography}%

\end{document}